\documentclass[preprint]{aastex}
\usepackage{xspace,graphicx}
\newcommand{\dor}{30~Doradus\xspace }

\newcommand{\co}[2]{$^{#1}$CO$\;${#2}}
\newcommand{\kms}{km$\;$s$^{-1}$}
\newcommand{\mjyb}{mJy beam$^{-1}$}
\newcommand{\jyb}{Jy beam$^{-1}$}

\newcommand{\centerclump}{52\xspace}
\newcommand{\eastclump}  {72\xspace}
\newcommand{\maserclump} {39\xspace}
\newcommand{\largepillar}{4\xspace}
\newcommand{\pillarN}    {21\xspace}
\newcommand{\pillarS}    {11\xspace}

\begin{document}

\title{ALMA Resolves 30~Doradus: Sub-parsec Molecular Cloud
  Structure Near the Closest Super-Star Cluster.}
\author{
R\'emy Indebetouw\altaffilmark{1,2},
Crystal Brogan\altaffilmark{1},
C.-H. Rosie Chen\altaffilmark{3},
Adam Leroy\altaffilmark{1},
Kelsey Johnson\altaffilmark{2},
Erik Muller\altaffilmark{4},
Suzanne Madden\altaffilmark{5},
Diane Cormier\altaffilmark{6},
Fr\'ed\'eric Galliano\altaffilmark{5},
Annie Hughes\altaffilmark{7},
Todd Hunter\altaffilmark{1},
Akiko Kawamura\altaffilmark{4},
Amanda Kepley\altaffilmark{1},
Vianney Lebouteiller\altaffilmark{5},
Margaret Meixner\altaffilmark{8},
Joana M. Oliveira\altaffilmark{9},
Toshikazu Onishi\altaffilmark{10},
Tatiana Vasyunina\altaffilmark{11}
}
\altaffiltext{1}{National Radio Astronomy Observatory, 520 Edgemont Road
Charlottesville, VA 22903; rindebet@nrao.edu, cbrogan@nrao.edu, aleroy@nrao.edu} 
\altaffiltext{2}{Department of Astronomy, University of Virginia, P.O. Box 3818, Charlottesville, VA 22903-0818; remy@virginia.edu}
\altaffiltext{3}{Max-Planck-Institut f\"ur Radioastronomie, Auf dem H\"ugel 69, D-53121, Bonn, Germany; rchen@mpifr-bonn.mpg.de}
\altaffiltext{4}{ALMA-J Project Office, National Astronomical Observatory of Japan, 2-21-1 Osawa, Mitaka, Tokyo 181-8588, Japan; erik.muller@nao.ac.jp, akiko.kawamura@nao.ac.jp}
\altaffiltext{5}{Service d’Astrophysique, Commissariat \'a L’\'Energie Atomique de Saclay, 91191 Gif-sur-Yvette, France; suzanne.madden@cea.fr, frederic.galliano@cea.fr}
\altaffiltext{6}{Institut f\"ur theoretische Astrophysik, 
Zentrum f\"ur Astronomie der Universit\"at Heidelberg, 
Albert-Ueberle Str. 2, D-69120 Heidelberg, Germany; diane.cormier@zah.uni-heidelberg.de}
\altaffiltext{7}{Max-Planck-Institut f\"ur Astronomie, K\"onigstuhl 17, D-69117, Heidelberg, Germany; hughes@mpia.de}
\altaffiltext{8}{Space Telescope Science Institute, 3700 San Martin Drive, Baltimore, MD 21218; meixner@stsci.edu}
\altaffiltext{9}{School of Physical and Geographical Sciences, Lennard-Jones Laboratories, Keele University, Staffordshire ST5 5BG, UK; j.oliveira@keele.ac.uk}
\altaffiltext{10}{Department of Physical Science, Graduate School of Science, Osaka Prefecture University, 1-1 Gakuen-cho, Naka-ku, Sakai, Osaka 599-8531, Japan; ohnishi@p.s.osakafu-u.ac.jp}
\altaffiltext{11}{Department of Chemistry, University of Virginia, Charlottesville, VA 22903}

\begin{abstract}
  
We present ALMA observations of 30~Doradus -- the highest resolution
view of molecular gas in an extragalactic star formation region to
date ($\sim$0.4pc$\times$0.6pc).  The 30Dor-10 cloud north of R136 was
mapped in \co{12}{2-1}, \co{13}{2-1}, C$^{18}$O$\;$2-1, 1.3$\;$mm
continuum, the H30$\alpha$ recombination line, and two H$_2$CO$\;$3-2
transitions.  Most $^{12}$CO emission is associated with small
filaments and clumps ($\lesssim$1pc, $\sim$10$^3\;$M$_\odot$ at the
current resolution).  Some clumps are associated with protostars,
including ``pillars of creation'' photoablated by intense radiation
from R136.  Emission from molecular clouds is often analyzed by
decomposition into approximately beam-sized clumps.  Such clumps in
30~Doradus follow similar trends in size, linewidth, and surface
density to Milky Way clumps. The 30~Doradus clumps have somewhat
larger linewidths for a given size than predicted by Larson's scaling
relation, consistent with pressure confinement.  They extend to higher
surface density at a given size and linewidth compared to clouds
studied at 10pc resolution.  These trends are also true of clumps in
Galactic infrared-dark clouds; higher resolution observations of both
environments are required.  Consistency of clump masses calculated
from dust continuum, CO, and the virial theorem reveals that the CO
abundance in 30~Doradus clumps is not significantly different from the
LMC mean, but the dust abundance may be reduced by $\sim$2.  There are
no strong trends in clump properties with distance from R136; dense
clumps are not strongly affected by the external radiation field, but
there is a modest trend towards lower dense clump filling fraction
deeper in the cloud.  
\end{abstract}

\section{Introduction}
\label{introduction}

\subsection{Star Formation and Feedback}

The evolution of a galaxy is determined by how efficiently and rapidly
it can turn interstellar gas into stars.  Several models have recently
been proposed to explain star formation rates and the interstellar gas
cycle in galaxies by self-regulating feedback
\citep{kmt09,oml10,hqm11,andrews11}. However, these models typically
only include a single dominant feedback mechanism among radiation
pressure, thermal/energetic input via an HII region, or mechanical
input from winds and supernovae.  All such models require more
detailed constraints from observations that resolve molecular clouds
near stellar clusters and associations that are energetically
affecting the interstellar medium (ISM).
In galaxies with sub-solar metal abundances, less efficient molecular
line and dust cooling may allow clumps to remain warmer and resist
gravitational collapse.  Reduced metallicity massive stars emit harder
radiation, and that radiation penetrates more deeply into molecular
clouds because of reduced dust abundances and possibly a clumpier
structure \citep[e.g.][]{poglitsch}.
To understand radiative transfer in molecular clouds, and the effect
of radiative feedback on star formation, one must resolve the
self-sheilded clumps from more diffuse inter-clump gas, in which CO is
more easily dissociated than H$_2$ \citep{wolfire10}.

Most star formation in a molecular cloud occurs in dense sub-parsec
sized clumps, which only constitute a fraction of the cloud's volume.
Detailed analysis of star formation physics thus requires sub-parsec
resolution and accurate column density and mass measurements, which
before 2012 was very challenging outside of the Milky Way.  However, a
much more extreme range of physical conditions is accessible outside
the Milky Way than within it: more massive stellar clusters with more energetic
feedback, metal abundances significantly different from solar, and
different galaxy-scale dynamics.  

The most extreme extragalactic molecular cloud and star formation
environments in the local universe are super-star clusters
(SSCs). SSCs form thousands of massive stars in a cubic parsec; 
young SSCs in the present-day universe offer a unique laboratory in
which to study the impact that this extreme mode of star formation has
on its environment, including subsequent star formation.  SSCs are
also likely a dominant mode during the era of galaxy assembly z$\sim$2
\citep{fallzhang01,johnson09}, so they are important to understand galaxy
evolution.

\subsection{Star Formation and Feedback at High Resolution in
  30~Doradus with ALMA}

A premier laboratory for studying feedback and star formation is
30~Doradus in the Large Magellanic Cloud (LMC).  At 50kpc
\citep{walker11}, R136 in 30~Doradus is the nearest SSC by an order of
magnitude.  R136 has $\gtrsim 10^6$ stars within a few parsecs,
allowing investigation of molecular cloud physics at reduced
metallicity \citep[$\sim$1/2 solar;][]{rolleston03,pagel03} bathed in
a strong radiation field.  Star formation in 30~Doradus has been going
on in multiple generations for $\sim$20Myr
\citep{demarchi11,walborn87}. By studying how new stars are forming in
this environment we can determine if the efficiency (stellar divided
by molecular mass) or stellar initial mass functions differ in SSC
environments.

CO has been studied previously in 30~Doradus, as part of LMC surveys,
notably \co{12}{1-0} with NANTEN \citep[][2.6\arcmin = 40pc]{fukui08}
and $^{12}$CO and \co{13}{1-0} with MOPRA \citep[33\arcsec = 8pc,
][]{wong11,hughes10}.  30~Doradus was among selected regions observed
with the Swedish-ESO Submillimetre Telescope \citep[SEST][and
references therein]{israel03}; \citet{johansson98} present \ $^{12}$CO
and \co{13}{1-0} at 45\arcsec=11pc resolution, and $^{12}$CO and
\co{13}{2-1} at 23\arcsec=6pc resolution.  Before the observations
presented here, the highest resolution observed in extragalactic CO
was 4pc \citep[\co{12}{3-2} in the LMC with SEST; ][]{johansson98}.

Spatial resolution limitations have resulted in a traditional divide
between extragalactic and Galactic studies of molecular gas.
Extragalactic studies usually consider giant molecular clouds (GMCs)
as entities 10-50$\;$pc in size, using a small number of abundant
molecules, and usually only the strongest optically thick transitions
of $^{12}$CO.  Populations of GMCs follow power-law scaling relations
in size, linewidth, and $^{12}$CO luminosity
\citep{larson,bolatto08,fukui08}.  By contrast, molecular gas emission
from Milky Way molecular clouds are routinely resolved with better
than 1pc resolution, and often better than 0.1pc.  Proximity makes it
possible to observe many transitions of multiple species. In
particular, for optically thick molecules like $^{12}$CO, a minor
isotopologue can be mapped to accurately determine column
densities.  Clumps on these smaller scales deviate from the
larger-scale scaling relations \citep{bp11,gibson09,ikeda09}.

The recently inaugurated Atacama Large (sub)Millimeter Array (ALMA
\footnote{This paper makes use of the following ALMA data:
  ADS/JAO.ALMA\#2011.0.00471.S. ALMA is a partnership of ESO
  (representing its member states), NSF (USA) and NINS (Japan),
  together with NRC (Canada) and NSC and ASIAA (Taiwan), in
  cooperation with the Republic of Chile. The Joint ALMA Observatory
  is operated by ESO, AUI/NRAO and NAOJ. The National Radio Astronomy
  Observatory is a facility of the National Science Foundation
  operated under cooperative agreement by Associated Universities,
  Inc.})  now allows extragalactic molecular clouds to be studied with
similarly high resolution and sensitivity as Galactic surveys,
revealing the detailed structure and physics.
This paper presents the first sub-parsec extragalactic investigation
of molecular emission, in 30~Doradus.
With ALMA we can address how molecular cloud structure is affected by
radiative feedback, and whether the dense star-forming structures in
clouds are different in such extreme environments.
The new ALMA observations are described in \S\ref{obs} as well as new
APEX observations and archival optical to mid-infrared data.  The
CO$\;$2-1 intensity distribution is analyzed in \S\ref{overview}, and
the relationship between CO emission and sources in the region is
discussed in \S\ref{ysos}.  In particular, for the first time we can
%associate dense molecular clumps with young protostars, and determine
%which protostars and young clusters have cleared their natal molecular
%gas.
% SUB2
isolate the denser parts of the molecular cloud directly involved with star formation from the lower density diffuse gas, 
and associate molecular clumps with young stellar and protostellar clusters.
% /SUB2
Section~\ref{mass} describes how temperature and mass are calculated
from CO and continuum emission.  Analysis of dense CO clumps is
presented in \S\ref{clumps}; comparison of clump masses calculated
from CO, dust, and the virial theorem constrains the molecular and
dust abundances, the amount of unobserved H$_2$ not traced by CO, and
the dynamical state of the clumps.  Key results are summarized in
\S\ref{conclusion}.

%XXX PdB,SMA,ATCA best linear work;  sensitivity a lot worse at least for ATCA.

%========================================================
\section{Observations}
\label{obs}
\subsection{ALMA}
\dor was observed with ALMA \citep{hills10} during Early Science
Cycle~0, using the 1.3$\;$mm Band 6 recievers \citep{kerr04,ediss04}.  The
spectral setup had 4$\times$1GHz FDM (frequency division mode) spectral
windows centered at 218.5, 219.267, 230.767, and 231.688GHz with
244kHz ($\sim$0.33km$\;$s$^{-1}$) resolution.  At the 250
km$\;$s$^{-1}$ LSRK velocity of \dor, these simultaneously cover
\co{12}{2-1} (230.538GHz rest frequency), \co{13}{2-1} (220.399GHz),
the H30$\alpha$ radio recombination line (231.90093GHz),
C$^{18}$O$\;$2-1 (219.560GHz),
% CCS 17-16  (220.21016GHz), 
H$_2$CO$\;3_{2,2}--2_{2,1}$ and $3_{2,1}--2_{2,0}$ (218.47563, 218.76007GHz), 
% 218.22219 is 3(0,3)-2(0,2) E_U=20.956K, other two have E_U=68K
HC$_3$N$\;$24-23 (218.32472GHz), and $^{13}$CS$\;$5-4 (231.2210GHz).
Observing dates and weather conditions are listed in Table~\ref{SB}.
In all observing sessions, the quasar J0538-440 was used as bandpass
calibrator (flux density $\sim$3Jy at the time of observation) and
J0635-7516 as phase calibrator (flux density $\sim$1Jy).  Callisto was
observed before the first observation on 2011-12-31, to set the
absolute flux scale with estimated uncertainty of $15\%$ using the
Butler-JPL Horizons-2010 model.  The flux scale was then transferred
to the other datasets by requiring the two quasars to have the same
flux densities.  The data were processed in the Common Astrononomy
Software Applications (CASA) package (\url{http://casa.nrao.edu}), and
visibilities imaged and deconvolved with 0.5$\;$km$\;$s$^{-1}$
channels.  The expected RMS per channel is 8$\;$mJy$\;$bm$^{-1}$, and
8--10$\;$mJy$\;$bm$^{-1}$ $\simeq$0.15$\;$K is achieved away from
strong lines (5--10 times worse sensitivity near bright lines because
the data there are dynamic range limited, i.e. bright emission creates
artifacts in the fainter parts of the image).
The synthesized beam is $\simeq$2.3\arcsec$\times$1.5\arcsec =
0.56pc$\times$0.36pc for \co{12}{2-1} and scales approximately with
frequency for other transitions.  The array configuration is sensitive to 
spatial scales from 1.1\arcsec to 20\arcsec (0.28 to 5$\,$pc).

\begin{deluxetable}{llllll}
\tabletypesize{\scriptsize}
\tablecolumns{6}
\tablewidth{0pc}
\tablecaption{Observations\label{SB}}
\tablehead {
execution block  & \# antennas & start & average   & PWV & time \\ 
    &     & time      & elevation &     & on source}
\startdata
uid://A002/X35edea/X228 & 13 & 2011-12-31 01:58:12 UT & 43$\deg$ & 3.5$\;$mm & 27.7min \\
uid://A002/X35edea/X5ad & 13 & 2011-12-31 05:00:33 UT & 40$\deg$ & 4.0$\;$mm & 27.7min \\
uid://A002/X35edea/X77d & 12 & 2011-12-31 06:31:23 UT & 33$\deg$ & 3.9$\;$mm & 27.8min \\
uid://A002/X36a70f/X895 & 15 & 2012-01-11 03:49:04 UT & 41$\deg$ & 2.2$\;$mm & 27.7min \\
\enddata
\end{deluxetable}

\subsection{APEX}

Observations with a single-dish telescope are necessary to recover the
total flux on large spatial scales, which is resolved out by an
interferometer.  Our single-dish observations were made in 2011
October 21--27, 29, 30 and December 3, 4, 12, and 13 with the 12~m
Atacama Pathfinder Experiment telescope, APEX\footnote{ APEX is a
  collaboration between the Max-Planck-Institut f\"ur Radioastronomie,
  the European Southern Observatory, and the Onsala Space
  Observatory.}  \citep{apex}.
APEX has a 23\arcsec\ beam at 230$\;$GHz.  Unfortunately, the ALMA
Compact Array was not yet available at the time of these observations.
The region was mapped twice using on-the-fly observations with the
APEX-1 single-sideband receiver that has a 2$\;$GHz spectral window
(consist of two 1$\;$GHz bands) and a full resolution of 30.5$\;$kHz
($\sim 0.0398$ km$\;$s$^{-1}$ at 230 GHz).  The mapping, centered on
(05:38:47.8, -69:04:50.3) with a size of 50\arcsec$\times$60\arcsec,
was first done in \co{13}{2-1} for 10h and for another 22h with the
spectral window covering \co{12}{2-1} and H30$\alpha$.  An
emission-free position at (05:40:19.40,-69:01:50.6) was used as the
off-source position and observed at the end of each scanning leg.  The
calibration was performed every $\sim$ 10 min with cold and ambient
temperature loads.
The ALMA correlator is more capable and permitted simultaneous
observation of additional lines (see below), but these are the only
three that we observed with APEX.

The data were calibrated with the APEX online calibrator (including
application of T$_A^*$ temperature scale) and then further reduced in
CLASS (\url{http://www.iram.fr/IRAMFR/GILDAS}) for baseline
subtraction on individual spectra and gridding.  The APEX data was
converted from main beam temperature to Jy$\;$bm$^{-1}$ by
multiplying by 39
\citep[][\url{http://www.apex-telescope.org/telescope/efficiency/}]{apex},
and the resulting cube was exported to CASA for combination with the
ALMA data.  
RMS noise was 0.5Jy/bm or 0.03K per 0.5$\;$\kms channel.
Peak intensities measured were 180$\;$\jyb = 4.7K and 21$\;$\jyb =
0.55K in \co{12}{2-1} and \co{13}{2-1}, respectively.
%We measure peak brightness temperatures of $^{12}$CO and \co{13}{2-1}
%of 70$\pm$2K and 19$\pm$1K, respectively, and peak integrated
%intensities of 445$\pm$20\kks and 105$\pm$10\kks.  
\co{12}{2-1} is
significantly brighter ($\times$3) than measured with SEST at similar
resolution (although more than an order of magnitude worse
sensitivity) to our APEX data \citep{johansson98}.  ALMA absolute flux
calibration is better than 10\%, and agrees with the APEX absolute
flux calibration to 20\%.  It is possible that the quoted SEST values
are not from exactly the same region -- since SEST does not have a data
archive, it is not possible to verify their findings.

The APEX image cube was combined with a cleaned ALMA cube using the CASA task
{\tt feather}, which Fourier transforms the two images, multiplies the single
dish image by the Fourier transform of the single dish beam, the interferometer
image by the complement of that transfer function, adds the two filtered
Fourier cubes, and transforms back. This feathered image is a simple way to
combine the two datasets but can be sensitive to the shape of the tapering
function and to relative calibration uncertainties.  To mitigate these effects
we used the feathered image as a model in clean deconvolution, which allows the
cleaning process to correct any potential issues with the feathered image which
might be inconsisent with the interferometric visibilities. 
Most (80\%) of the APEX flux is recovered by ALMA, so the properties
of small clumps do not depend strongly on how the datasets are
combined. (In this paper ``clump'' refers to the structures containing
local maxima and $\sim$0.5-2pc in size, $\sim$100-1000$\;$M$_\odot$ in
mass, defined by the decomposition described in \S\ref{clumpdecomp}).
We experimented with small variations in the relative flux calibration
between the two telescopes, and conclude that the image combination
process results in less than 10\% uncertainty of $^{12}$CO clump
properties and less than 5\% uncertainty of $^{13}$CO clump
properties.

\subsection{Archival Data}
\label{archivalsection}
Multiwavelength data was retrieved from various archives as listed in
Table~\ref{archival}.  We processed ESO data through the {\tt gasgano}
data pipeline (\url{http://www.eso.org/sci/software/pipelines/});
other telescopes provide calibrated reduced archival images.
Astrometric registration was performed iteratively starting with 2MASS
and {\it Spitzer}/SAGE, then ground-based images, followed by HST,
which has the smallest field of view.  Final image registration and
distortions are estimated to be $\lesssim$0.4$\arcsec$ over the ALMA
observed region.  Near-infrared (NIR) data were cross-calibrated to
the HST flux scale, then background-matched and mosaicked using
Montage (\url{http://montage.ipac.caltech.edu/}).  Complete
coverage at the highest resolution is not available in all filters; in
particular for the NIR images discussed below, the F160W and F110W
images are from HST, whereas the F205W image is largely from
ground-based SOFI, with NICMOS data inset where available.

\begin{deluxetable}{lllllll}
\tabletypesize{\scriptsize}
\tablecaption{\label{archival}Archival data used in this study.}
\tablehead{
instrument & filter/band & resolution & project ID & PI & date & reference }
\startdata
WFC3\tablenotemark{1}  & F110W,F160W $\sim$J,H & 0.15\arcsec& 12939 & Sabbi & 2012 & \\
NICMOS& F205W $\sim$K$_s$ & 0.2\arcsec & 7819 & Walborn & 1997-8 & \citet{walborn99}\\ 
SOFI  & J,K$_s$   & 0.75\arcsec & 63.H-0683(A) & Schmutz & 1998-9 & \citet{crowther02}\\
Spitzer\tablenotemark{2}/SAGE 
& IRAC 3.5-8$\mu$m & 2\arcsec & 20203 & Meixner & 2005-6 & \citet{meixner06} \\
ATCA\tablenotemark{3}
& 3cm,6cm & 2\arcsec & C109,C234 & Lazendic, Dickey & 1993,98,99 & \citet{lazendic_cm} \\
\enddata
\tablenotetext{1}{Observations made with the NASA/ESA Hubble Space Telescope, obtained from the data archive at the Space Telescope Science Institute. STScI is operated by the Association of Universities for Research in Astronomy, Inc. under NASA contract NAS 5-26555}
\tablenotetext{2}{Spitzer Space Telescope is operated by the Jet Propulsion Laboratory, CalTech under a contract with NASA}
\tablenotetext{3}{The ATCA is part of the Australia Telescope funded by the Commonwealth of Australia for operation as a National Facility, managed by CSIRO}
\end{deluxetable}

%\clearpage
%========================================================
\section{Results: High-Resolution Millimeter Emission in 30 Doradus}
\label{overview}

For the first time we can study the detailed subparsec-scale clump
mass distribution in extragalactic clouds.  In Cycle~0, ALMA had 1/4
of its expected final sensitivity (and more than 30 times coarser
spatial resolution), so the observations presented here were designed
as a pilot program.  At $>$10pc resolution, there are two GMCs within
20pc of the R136 cluster; we chose to observe the northern, more
massive cloud 30Dor-10 \citep[nomenclature from][]{johansson98}.  In
the following subsections we present the 3-dimensional CO emission
distribution, describe other detected lines and the continuum.
Subsequent sections will connect the CO emission to current star
formation detected with infrared and maser emission, and
quantitatively analyze the molecular mass distribution and clump
properties.

%===================================================

% rms 8mJy/bm in 0.5 km/s chan cube;  
\begin{figure}[h!]
%\centerline{\resizebox{4in}{!}{\includegraphics{30Dor_midIR_mol_1panel.eps}}}
\centerline{\resizebox{4in}{!}{\includegraphics{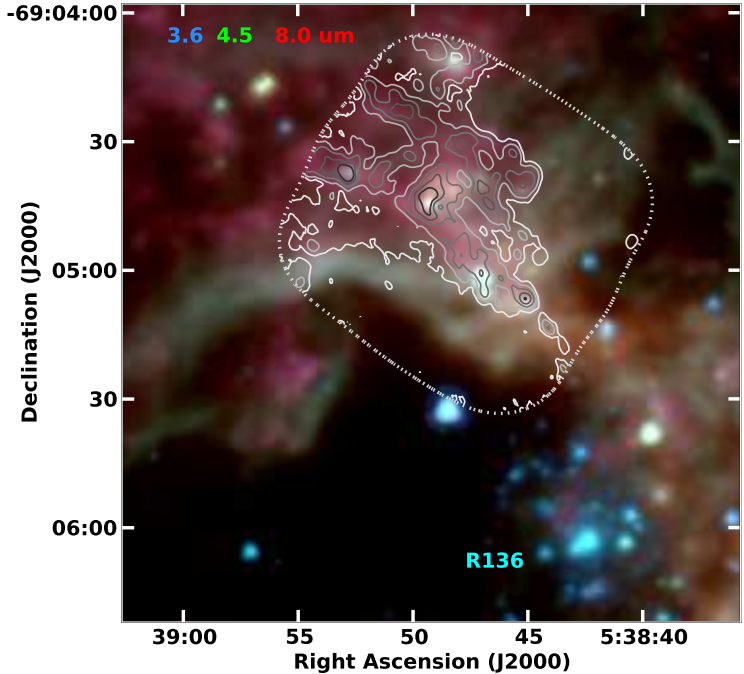}}}
\caption{ SAGE {\it Spitzer} IRAC 3-color image of the northern
part of 30~Doradus.  Red, green, and blue are 7.9, 5.8, and
3.6$\mu$m, respectively.  $^{12}$CO$\;$2-1 integrated intensity is overlaid with contour levels 2, 6, 15, 30, 45 \jyb\/\kms; RMS noise is 0.2 \jyb\/kms.
A dashed contour at the 20$\%$ FWHP point of the ALMA mosaic is also shown.  
R136 is the blue cluster located 45\arcsec\xspace ($\simeq$11pc) to the south of the field of view mapped by ALMA. 
\label{overviewfig}}
\end{figure}

\begin{figure}[h!]
%\centerline{\resizebox{6in}{!}{\includegraphics[trim=0 250 0 0,clip=true]{30Dor_12CO_mom0.eps}}}
%\centerline{\resizebox{6in}{!}{\includegraphics[trim=0 250 0 0,clip=true]{f2.eps}}}
\centerline{\resizebox{6in}{!}{\includegraphics{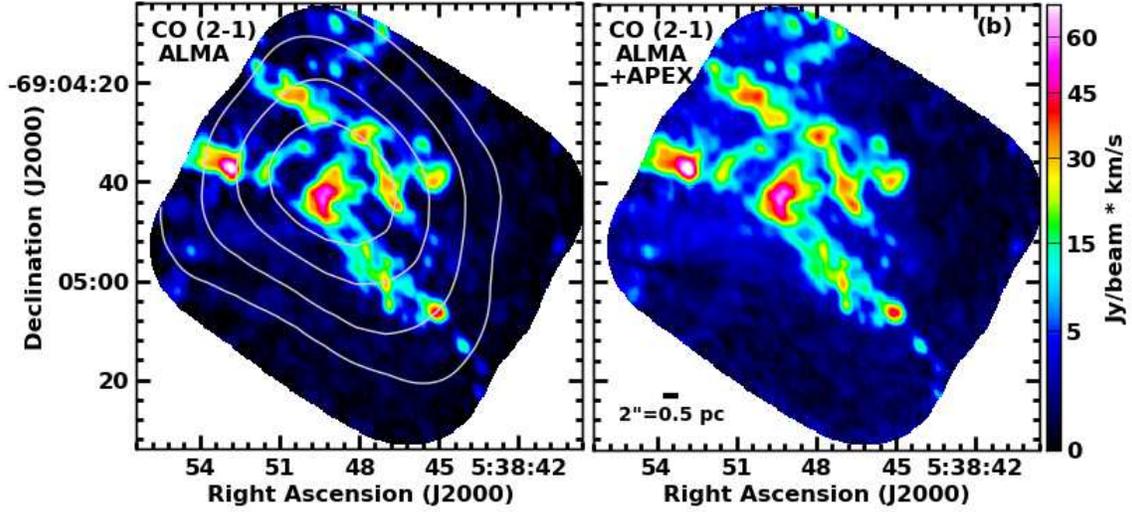}}}
\caption{(a) ALMA-only integrated \co{12}{2-1} intensity image,
  overlaid with contours of APEX intensity. (b) combined ALMA+APEX
  \co{12}{2-1} integrated intensity. \label{apexcombine}}
\end{figure}

\subsection{$^{12}$CO and \co{13}{2-1} Emission}

\begin{figure}[h!]
%\resizebox{3.5in}{!}{\includegraphics{30Dor_13CO_mom0.eps}}
\resizebox{3.5in}{!}{\includegraphics{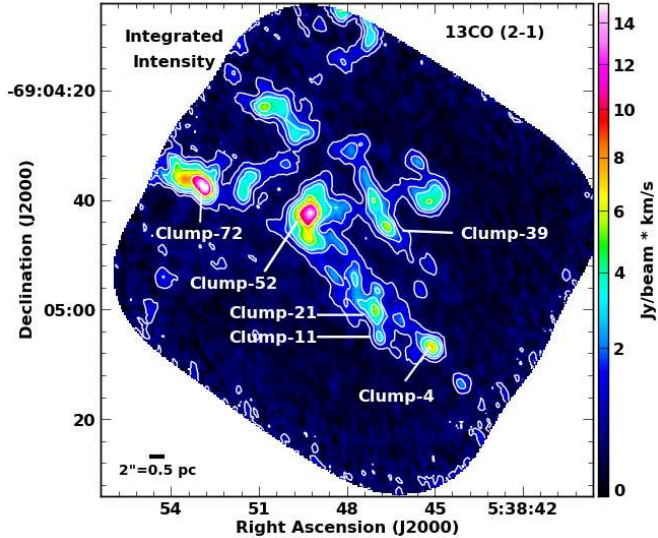}}
\caption{\label{int13co} Image of the ALMA+APEX \co{13}{2-1}
  integrated intensity in 30Dor-10. The contour levels are 0.6, 1.6,
  2.6, 4.6, and 8.6 \jyb\/\kms\/. Notable CO clumps are labeled for
  reference.}
\end{figure}

Figure~\ref{overviewfig} shows integrated \co{12}{2-1} emission
together with mid-infrared continuum \citep[from the {\em
  Spitzer}/IRAC SAGE program, ][]{meixner06}.  For the first time, the
molecular gas is imaged at the same resolution as {\em Spitzer}/IRAC.
$^{12}$CO is detected across most of the mapped region.  The R136
cluster is 11~pc in projection southwest from the bottom of 30Dor-10.
The edge of the bubble evacuated by R136 is clear in both infrared and
CO emission -- the high emission measure edge is enhanced at 4.5$\mu$m
(green) because of bright Br$\alpha$ recombination line emission in
that IRAC band.  Deeper in the molecular cloud, knots of CO emission
correlate to some extent with MIR continuum from protostars; detailed
comparison will be discussed in \S\ref{ysos}.  The \co{12}{2-1}
integrated maps using ALMA and APEX alone are compared to the combined
map in Figure~\ref{apexcombine}.
Figure~\ref{int13co} shows integrated \co{13}{2-1} emission, noting
the numbers of the most prominent clumps (clump decomposition will be
described in section~\ref{clumpdecomp}).  

\begin{figure}[h]
%\resizebox{3.5in}{!}{\includegraphics{intenshist.w3.eps}}
\resizebox{3.5in}{!}{\includegraphics{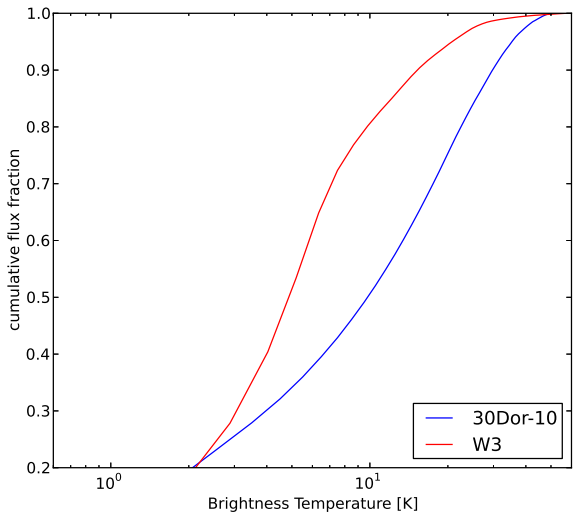}}
\caption{\label{intenshist} Comparison of the cumulative $^{12}$CO
  intensity distribution in 30Dor-10 and in the Galactic massive star
  formation region W3.  Both datasets have similar physical
  resolution, peak temperature, and brightness temperature
  sensitivities \citep{bieging}.  A much larger fraction of W3 has
  low surface brightness than 30Dor-10.}
\end{figure}

There is not an overwhelming amount of large-scale or extended
$^{12}$CO emission in 30Dor-10.  As already noted, 80\% of the APEX
flux is recovered by ALMA. The intensity distribution is further
quantified in Figure~\ref{intenshist} which compares the cumulative
brightness distribution in 30Dor-10 and in the Galactic massive star
formation region W3 \citep{bieging}.  The W3 \co{12}{2-1} cube has
physical resolution of 0.3pc (at 2kpc distance), 0.15K RMS noise per
0.5$\;$\kms channel, and peak antenna temperature of 54K, compared to
the same quantities in 30Dor-10 of 0.5pc, 0.15K, and 62K, respectively
(\S\ref{obs}).  There is more diffuse $^{12}$CO emission in a Galactic
region like W3 than in 30Dor-10.  A reasonable interpretation of this
result is that at subsolar metallicity and increased radiation
density, the less-dense ``interclump'' CO is being selectively
photodissociated as expected from PDR models \citep{wolfire10}.

% SUB2
Models of low-spatial-resolution molecular line emission derive volume
and area filling fractions in 30Dor-10 using escape probability
techniques.  Derived area filling fractions using CO$\;$1-0 to
CO$\;$3-2 include 0.05-0.1 \citep{johansson98}, 0.05 \citep{heikkila},
0.015 \citep{nikolic07}.  In our integrated intensity maps, 2/3 of the
\co{12}{2-1} flux is contained in 15\% of the spatial area, and half
of the flux in 8.9\% of the spatial area.  For \co{13}{2-1}, 2/3 of
the flux is contained in 9.8\% of the area, and half of the flux in
5.6\% of the area.  Thus most of the flux in our map is contained
within 5--10\% of the area, consistent with the escape
probability-derived filling fractions.  \citet{pineda12} derived 0.06
from CO$\;$4-3 and CO$\;$7-6 -- these higher J transitions should be
less sensitive to optical depth effects and a better tracer of the
dense gas, but also may not be the same structures emitting CO$\;$2-1.
% /SUB2

\subsection{CO Velocity Structure}

\begin{figure}[h!]
\centerline{
%\resizebox{!}{2.7in}{\includegraphics[trim=60pt 50pt 20pt 50pt,clip]{13CO.demom0.2.eps}}
%\resizebox{!}{2.7in}{\includegraphics[trim= 5pt 50pt 20pt 50pt,clip]{13CO.ramom0.2.eps}}}
\resizebox{!}{2.7in}{\includegraphics[trim=60pt 50pt 20pt 50pt,clip]{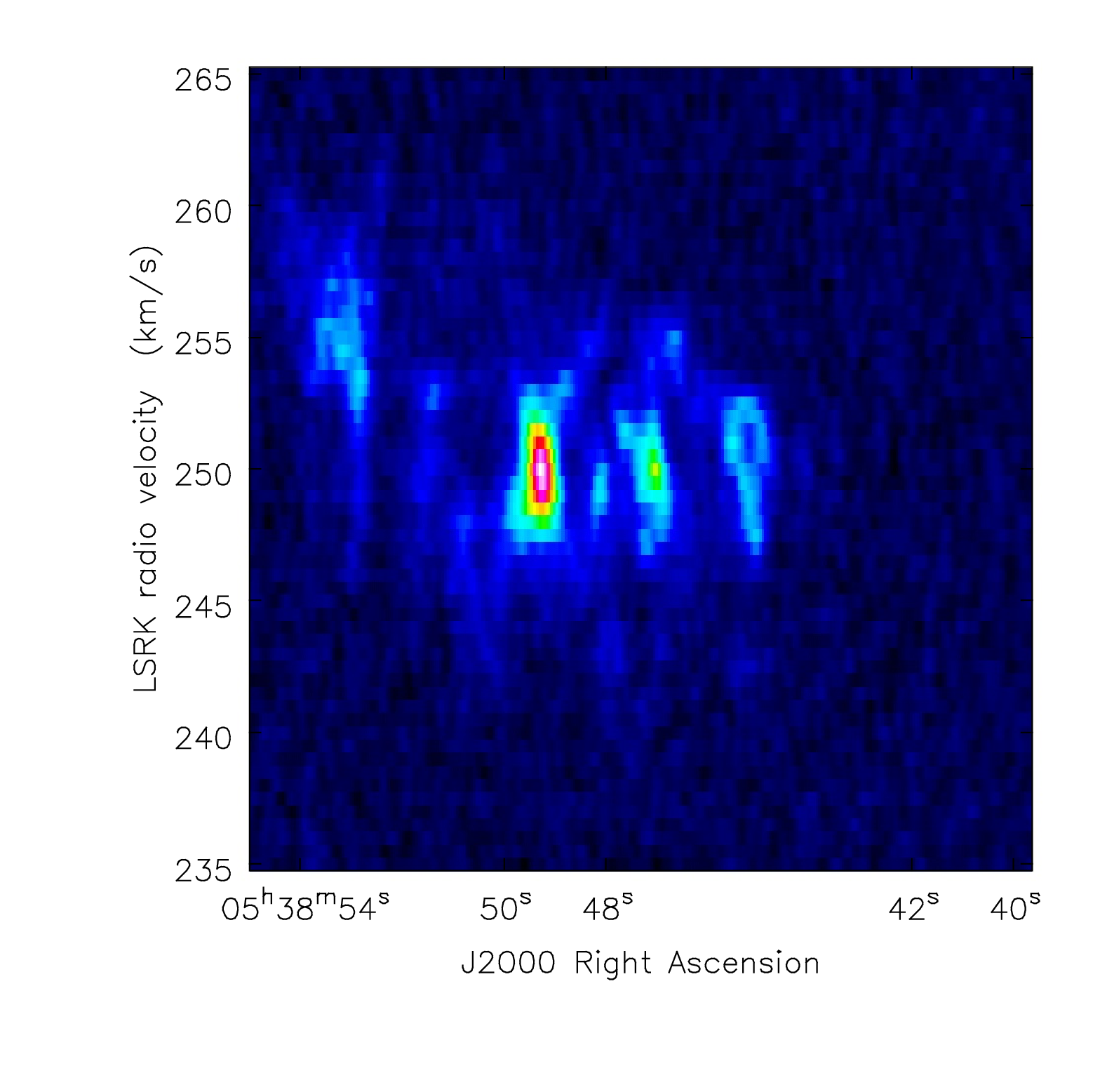}}
\resizebox{!}{2.7in}{\includegraphics[trim= 5pt 50pt 20pt 50pt,clip]{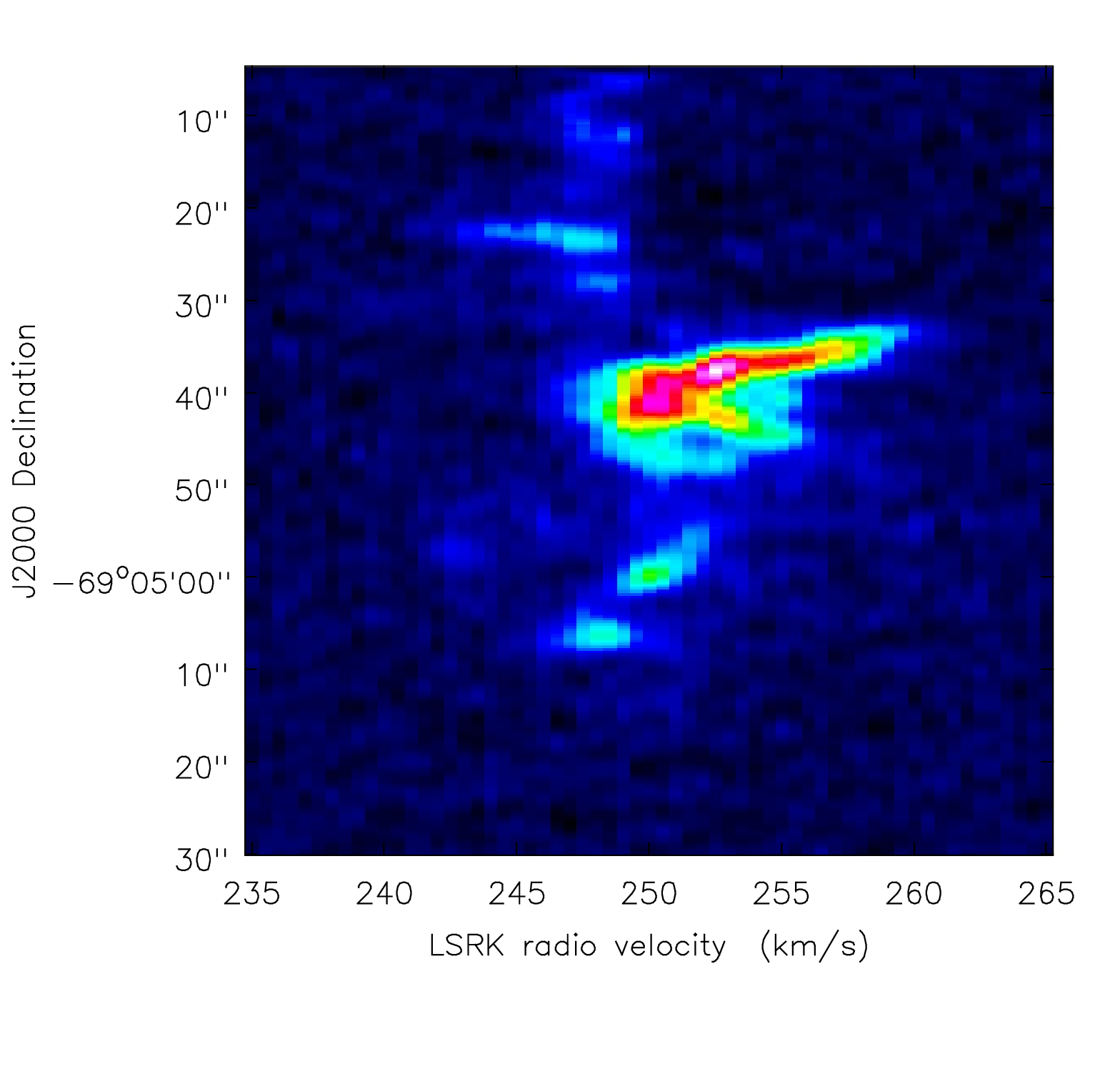}}}
\caption{Integrated intensity images of the \co{13}{2-1} collapsed along each spatial axis of the cube, i.e. intensity as a function of Right Ascension (R.A.) and velocity and of Declination and velocity.\label{moment0}}
\end{figure}

Figure~\ref{moment0} shows integrated $^{13}$CO emission as a function
of right ascension (R.A.) and velocity, and as a function of
declination and velocity, i.e. the cube collapsed along each spatial
axis.  The cloud is highly structured in position and velocity - in
particular the brightest clump \centerclump near the center of
the map may be at the intersection of two converging flows, as
expected from simulations of turbulent clouds
\citep[e.g.][]{gong11,heitsch11}.

\begin{figure}[h!]
%\resizebox{6in}{!}{\includegraphics{30Dor_moments.eps}}
\resizebox{6in}{!}{\includegraphics{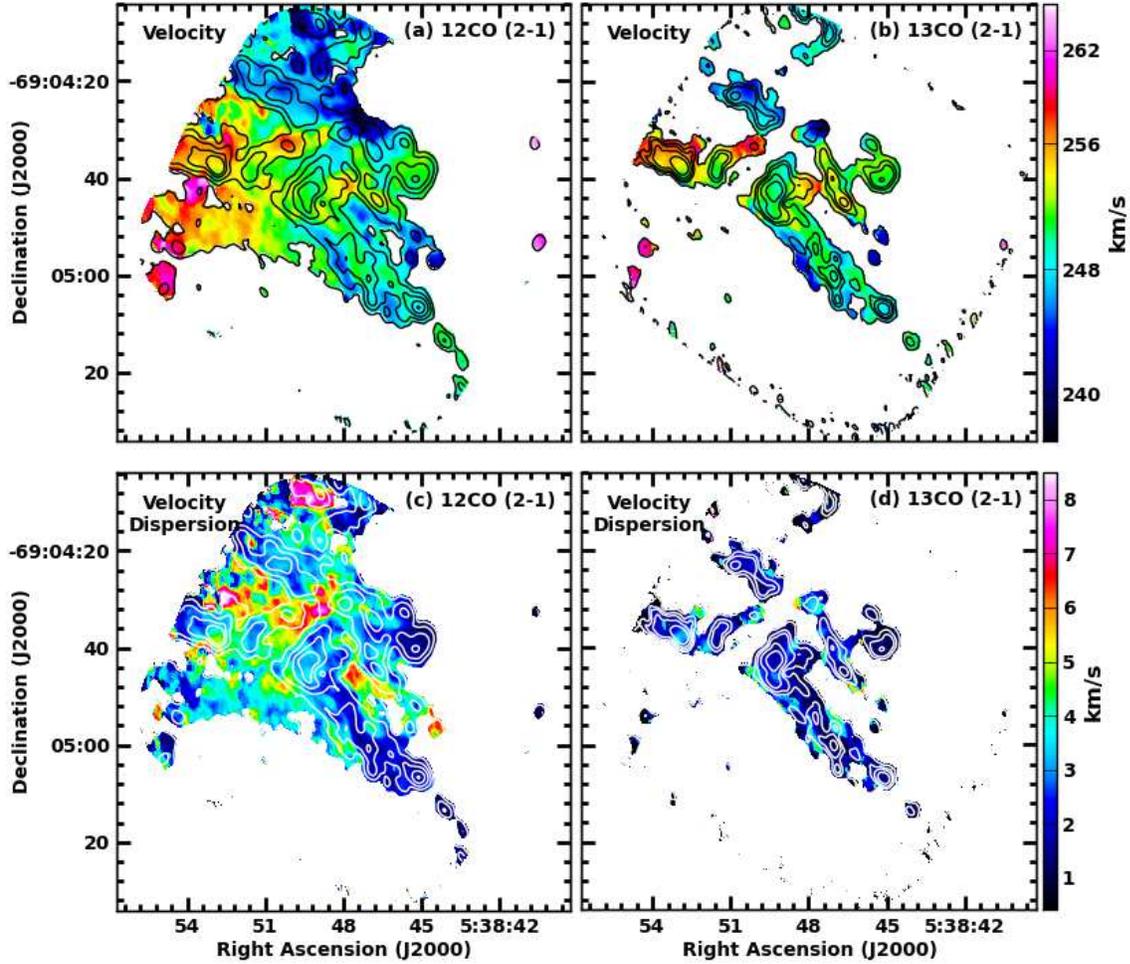}}
\caption{\label{moments}
Images of the ALMA+APEX first moment (mean velocity) for (a) $^{12}$CO$\;$2-1 and (b) $^{13}$CO$\;$2-1, along with images of the second moment
(velocity dispersion) for (c) $^{12}$CO$\;$2-1 and (d) $^{13}$CO$\;$2-1.
Integrated intensity contours of $^{12}$CO$\;$2-1 at are shown in (a)
and (c) at 2, 6, 15, 30, and 45 \jyb\/\kms\/. Integrated intensity
contours of $^{13}$CO$\;$2-1 at are shown in (b) and (d) at 0.6, 1.6,
2.6, 4.6, and 8.6 \jyb\/\kms\/.}
\end{figure}

The overall velocity structure of the cloud has a moderate east/west
gradient, most evident in $^{12}$CO emission, although the much more
dominant features are the relative velocities of individual clumps,
typically separated by 3-10$\;$\kms.  Figure~\ref{moments} shows the
first and second velocity moments.  Clump linewidths are 2--3$\;$\kms\
FWHM (see analysis in section \ref{size-linewidth} below).  The second
velocity moment (velocity dispersion) shows a few high peaks
(Figure~\ref{moments}), but most of those are merely the spatial
superposition of two clumps at different velocities.  Clump~\eastclump
truly does have a very large linewidth (10.5$\;$\kms in \co{12}{2-1}).
That clump contains an embedded mid-IR protostar or protocluster (see
Section~\ref{ysos} below), and the large linewidth likely results from
massive outflows.

\clearpage
%=================================================================
\subsection{C$^{18}$O, H$_2$CO and Other Lines}

\begin{figure}[h!]
%\resizebox{3.5in}{!}{\includegraphics{30Dor_C18O_mom0.eps}}
%\resizebox{3in}{!}{\includegraphics{core0.eps}}
\resizebox{3.5in}{!}{\includegraphics{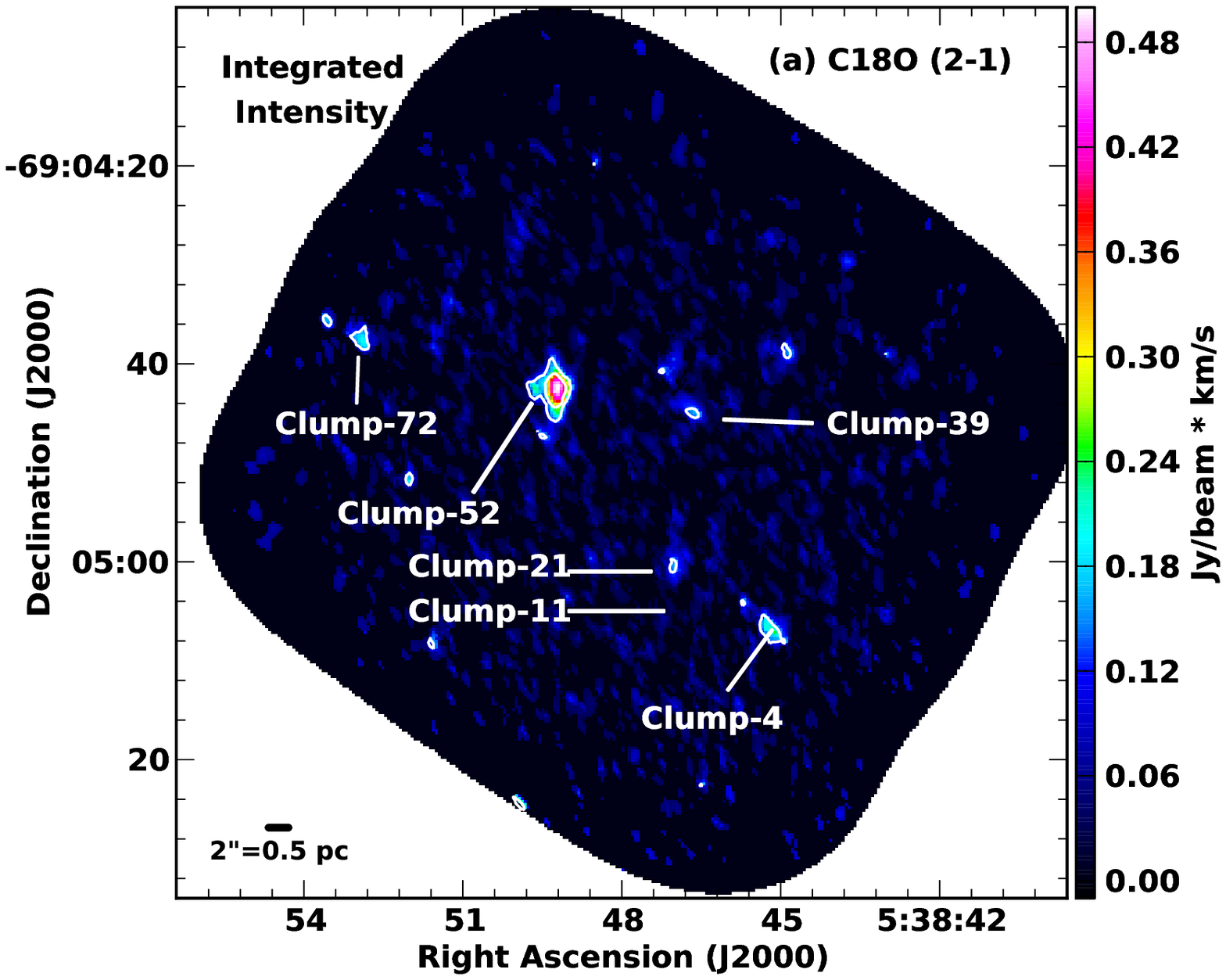}}
\resizebox{3in}{!}{\includegraphics{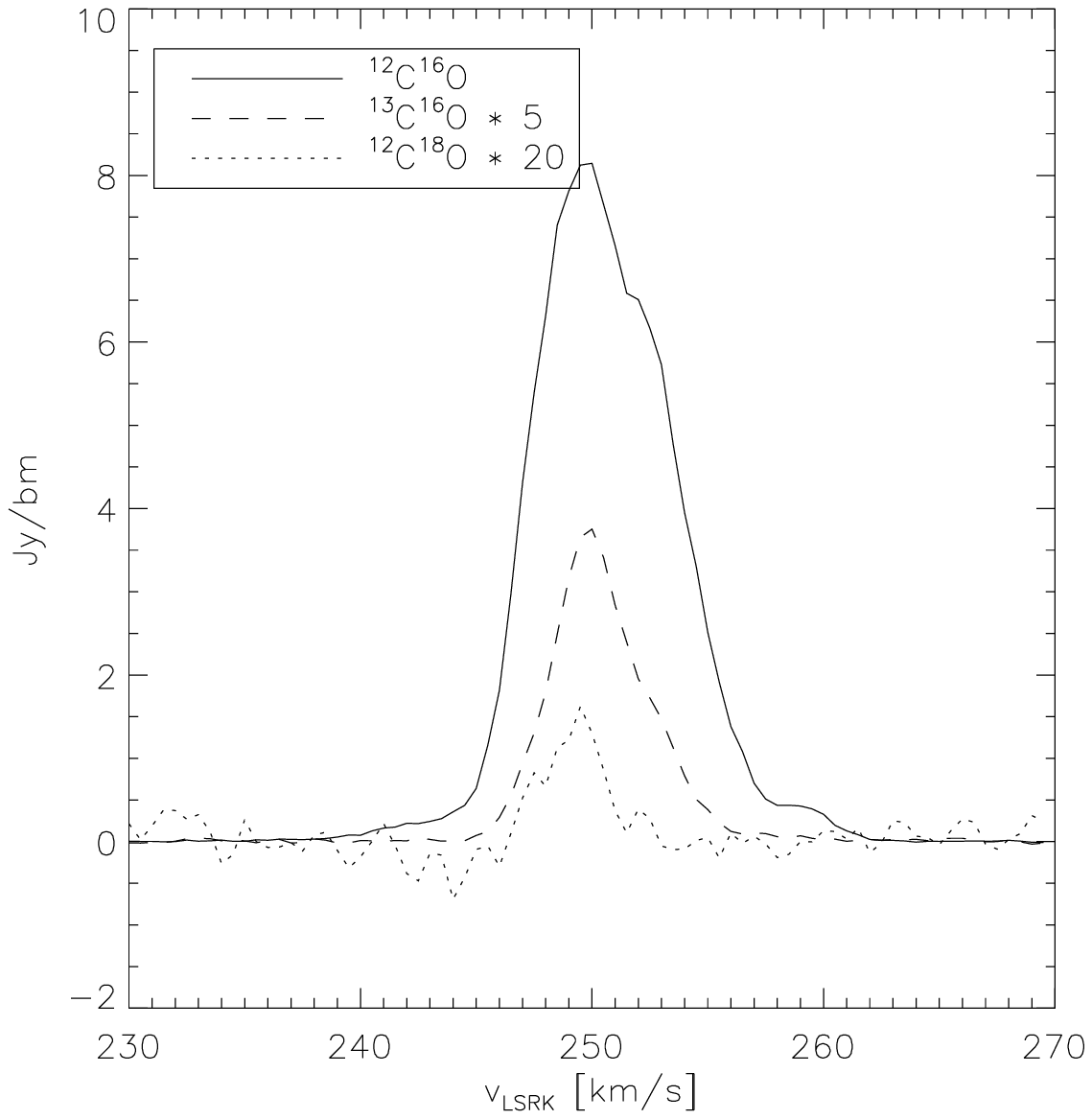}}
\caption{(a) Image of the C$^{18}$O (2-1) integrated intensity. The
  contour levels are 0.14 and 0.28 \jyb\/\kms\/. (b) Spectra of all
  three CO isotopologues toward the brightest
  clump~\centerclump.\label{c18o}}
\end{figure}

Several other lines were observed simultaneously with \co{12}{2-1} and
\co{13}{1-0}.  C$^{18}$O is detected in the brightest central clump
\centerclump, the tip of the pillar/clump \largepillar, as well as
several other locations in the cloud (Figure~\ref{c18o}).
H$_2$CO 3-2, HC$_3$N 24-23,
and $^{13}$CS 5-4 are not convincingly detected, although
there are some suggestive spectral features associated with the
brightest $^{13}$CO peaks at the $\sim$2$\sigma$  level.
HC$_3$N nondetection is consistent with previous single-dish observations, 
but H$_2$CO nondetection may be a selection effect, since due to correlator restrictions we observed the weakest of three 3-2 lines, and astrochemical models predict fairly strong H$_2$CO \citep{millar90}.
We also detected the H30$\alpha$ radio recombination line (RRL;
$\nu_{rest}$ = 231.90093$\;$GHz) in 30~Doradus. The RRL emission is
mostly coincident with peaks in the 3cm radio continuum emission in
\citet{lazendic_cm} and young massive stars in \citet{parker93}, and
not coincident with molecular gas peaks.  Analysis of this data will
be presented in Kepley et al. (2014, in prep).
HCN$\;$1-0, HCO$^+\;$1-0, CS$\;$2-1, and H40$\alpha$ at $\sim$90GHz
were also observed as part of this project and will be presented in
Brogan et al. (2013, in prep).
Note that we do not have single-dish data for any of these lines except
for H30$\alpha$, which is not detected with APEX.

\subsection{Continuum}
\label{continuumdata}

\begin{figure}[h!]
%\resizebox{!}{3.in}{\includegraphics{30Dor_continuum.eps}}
\resizebox{!}{3.in}{\includegraphics{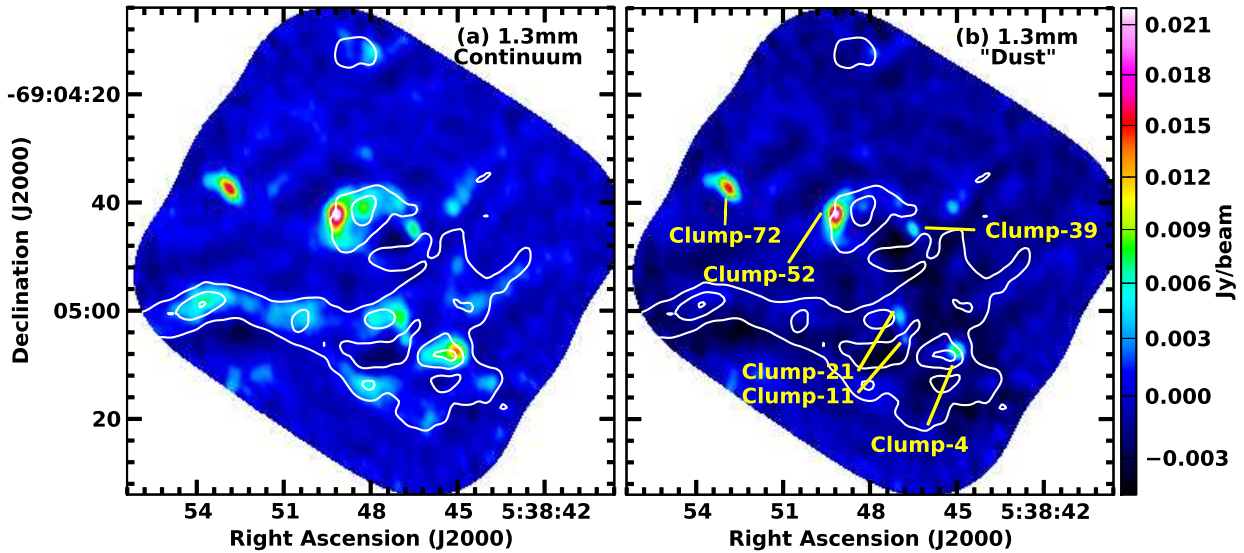}}
\caption{(a): 1.3$\;$mm continuum image and 6~cm continuum contours at
  4, 8, 12, and 16 \mjyb\/ (Lazendic et al. 2003). (b): ``Dust-only''
  emission map constructed by subtracting scaled free-free emission
  (6~cm) from the 1.3$\;$mm image, again with the 6~cm contours overlaid.
 Several of the $^{13}$CO clumps are labeled for reference.
  \label{cont}}
\end{figure}

The 1.3$\;$mm continuum is also imaged with high fidelity throughout the
region.  Note that unlike the CO cubes analyzed throughout this paper,
we do not have single-dish continuum observations, and some
large-scale diffuse emission may be resolved out (the expected largest
angular size from the uv distribution of the data is 20\arcsec).
Figure~\ref{cont} compares the ALMA 1.3$\;$mm continuum to 6$\;$cm continuum
imaged at similar resolution with the Australia Telescope Compact
Array \citep{lazendic_cm}.  Those data have a synthesized beamsize of
1.83\arcsec$\times$1.75\arcsec, and no single dish data included.  The 1.3$\;$mm
continuum traces the rim on the southern part of the cloud, and likely
includes a significant contribution from free-free emission at that
location, but further north is morphologically distinct from tracers
of ionized gas, and is thus likely mostly dust emission.  In
particular the central and eastern dense gas clumps \centerclump and
\eastclump are bright in the continuum and both contain an embedded
protostar or protocluster (see Section~\ref{ysos}).

The free-free contribution to the 1.3$\;$mm continuum can be
subtracted by assuming that the 6$\;$cm continuum is 100\% optically thin
free-free emission, and scaling S$_\nu$ by $\nu^{-0.1}$.  To properly
scale and subtract the 6$\;$cm image from the 1.3$\;$mm image requires
matching the {\it uv} coverage of the two interferometers.  Although
the restoring beams are not drastically different, a simple scaling by
the beam areas (the images have native units of Jy$\;$bm$^{-1}$) and
$\nu^{-0.1}$ factor results in negative values over significant parts
of the ``dust emission'' map.  To better perform the subtraction, we
spatiallly filtered the 6$\;$cm image to match it to ALMA's spatial
sensitivity, and then convolved both maps to 2.67\arcsec$\times$2.0\arcsec.
A total of 39\% of the total flux in the 1.3$\;$mm image is removed as free-free
contamination.  This may still represent oversubtraction of the
free-free contribution, but we choose to be conservative in what we
maintain as dust emission in subsequent analysis.  The millimeter flux
densities of clumps in the body of the molecular cloud are insensitive
to this scaling, since their free-free contribution is much smaller
than the dust (the cm flux density is much less than the mm flux
density).  Those on or near the photodissociated rim including
clump~\largepillar, \pillarN, and \pillarS, have a larger free-free
contribution.  We also performed the free-free subtraction using the
3$\;$cm image from \citet{lazendic_cm}, with quantitatively similar
results.

%\clearpage
% ==============================================================
\section{The Relationship Between Molecular Emission and Current Star Formation}
\label{ysos}

\begin{figure}[h!]
%\resizebox{6in}{!}{\includegraphics{30Dor_JHKs_mol_zoom.eps}}
\resizebox{6in}{!}{\includegraphics{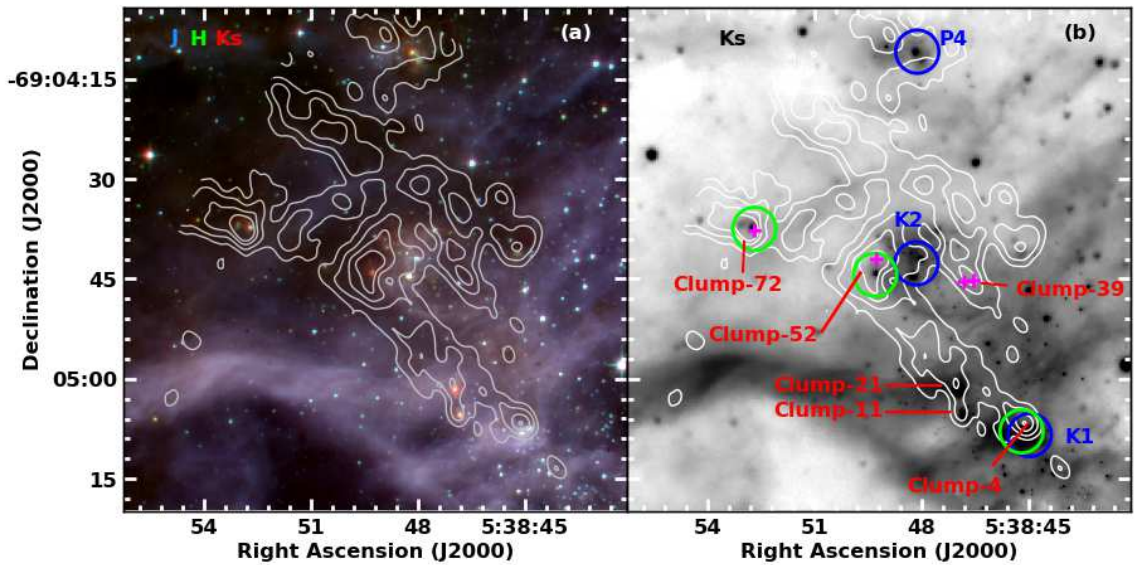}}
\caption{(a): Near-IR JHK$_s$ mosaic with \co{12}{2-1} integrated
  intensity contours.  (b): K$_s$ image with annotations and
  \co{12}{2-1} contours. NIR clusters are indicated by blue circles, {\it
    Spitzer} identified massive young stellar object candidates have
  green circles, and the locations of 
H$_2$O masers are indicated by magenta $+$ symbols (see text for
references).  Notable CO clumps are also labeled in red. In both
panels, the \co{12}{2-1} integrated intensity contour levels are
6, 15, 30, 45 \jyb\/\kms\/.
\label{nir_ysos}}
\end{figure}

Figure~\ref{nir_ysos} shows subarcsecond resolution near-infrared
(NIR) images (mostly from HST; see \S\ref{archivalsection}) overlaid
with \co{12}{2-1} contours and with previously known sources marked as
well as notable CO clumps.  Of particular interest is the relation
between dense molecular clumps and star formation.  We discuss
previously identified star formation from most to least evolved,
namely NIR protostellar clusters, mid-infrared massive YSOs, and
masers.  

The compact cluster K1 (original designation from \citet{walborn87},
studied in more detail by \citet{rubio98,walborn99}) is located at the
tip of pillar/clump \largepillar, in the classic ``pillars of
creation'' arrangement seen throughout the Milky Way
\citep{hester}. The stars are associated with diffuse NIR emission, likely
Brackett$\gamma$ emission from the ionized gas being photoablated from
the pillar tip.
To the east are two very red young stellar objects only marginally
resolved at HST resolution \citep[IRSN122 and 126 in][]{rubio98}.
Both YSOs are also located in a pillar-like structure (clumps \pillarN and
\pillarS).  These three regions are morphologically suggestive of
triggered star formation via radiatively driven implosion
\citep{sandford82,bertoldi89}, but {\em proving} triggering is
complex, and these may simply be {\em revealed} clusters that formed
in the relatively denser molecular clumps, now left behind after
less-dense outer layers are ablated by radiation from R136 to the
south \citep[e.g.][]{dale07}.

Cluster K1 is unresolved by {\it Spitzer} IRAC, and was designated a
massive young stellar object (MYSO) candidate by \citet{gc09}, based
primarily on its 5.8$\mu$m-8.0$\mu$m color and infrared morphology. It
is somewhat extended in IRAC images, and thus not included in the more
conservative SAGE point source catalog \citep{meixner06} or MYSO list
derived from it \citep{whitney08}. It is not clear that the cluster
resolved by HST contains any intermediate or massive embedded YSOs.  
% SUB2
This example does highlight a general issue with {\it Spitzer} point sources, which we and the identifying authors all call MYSOs:  these are almost certainly small clusters or multiples.  The size and mass scales ($\sim$10$^3\;$M$_\odot$, $\lesssim\;$1pc) probed by the current ALMA data are not small enough to be associated with individual protostars, but on the other hand it is unlikely that any massive star formation is {\it not} associated with a clump described here.
%/SUB2

CO clump \largepillar immediately behind (in projection, from the
point of view of R136) K1 is however a likely candidate to contain a
very young MYSO.  Clump \largepillar has $^{12}$CO and \co{13}{2-1}
peak brightness temperatures of 61K and 13K respectively and is very
compact, suggesting that it may be internally heated (although no IRAC
source can be separated from K1 and other diffuse emission in the
area).
OH maser emission at 1720MHz was detected coincident with clump
\largepillar and cluster K1 (circle in Fig.~\ref{nir_ysos}) by
\citet{brogan04} and \citet{roberts05}.  1720MHz OH masers are
typically found toward both star formation regions and supernova
remnants; in this case the maser velocity 243$\;$km$\;$s$^{-1}$ is
offset from the clump velocity 248$\;$km$\;$s$^{-1}$, suggesting that
the emission is {\em not} associated with the star formation, but on
the other hand, previous searches for supernova remnants within \dor
have also been unsuccessful or inconclusive \citep{chu04}.

The cluster K2 in the center of the field is not cospatial with a CO
clump -- the nearest (clump \centerclump) is adjacent.  Presumably,
the cluster has largely dispersed the dense gas from which it formed,
although there are remnants visible in the irregular filamentary
structure extending from CO clump~\centerclump to NIR cluster K2.  A
greater spread in age and/or reddening in the optical color-magnitude
diagram of K2 presented by \citet{walborn02} supports the notion that
it may be more evolved than K1.  The cluster has also created its own
small HII region, evident in the cm continuum (\S\ref{continuumdata}).
A high \ion{S}{4}/\ion{S}{3} ratio \citep{ri_dor} indicates that this
HII region is more highly ionized (higher ionization parameter and/or
harder ionizing field) than the ionized gas on the rim of the bubble
around R136 (the ridge at the bottom of our field).  \citet{walborn13}
classify the brightest source in the cluster, Parker~1429, as
O2V((f*))z; this very young and hot star may be the source of
relatively hard ionizing radiation.

The CO clumps on either side of cluster K2 each contain much younger
MYSOs.  Clump~\maserclump to the west contains a 22.235GHz H$_2$O
6$_{16}$-5$_{23}$ maser at the same velocity as the CO emission
\citep[pink $+$ on
Figure~\ref{nir_ysos}][]{whiteoak83,oliveira_maser,lazendic_maser,imai}.
In the Milky Way, water masers are commonly associated with massive
young stellar objects \citep[e.g.][]{churchwell90}, and they are
beginning to be used in more distant galaxies as signposts of massive
star formation as well \citep[e.g.][]{darling08,brogan10}.  The masers
require densities $\gtrsim$10$^9\;$cm$^{-3}$ and warm gas
$\gtrsim$400$\;$K, and are formed in outflows or compressed
circumstellar material around MYSOs.  In the Galaxy, even MYSOs that
are sufficiently embedded that they do not have detectable NIR
emission are sometimes associated with 22GHz H$_2$O masers; there is
no detectable MIR emission in clump~\maserclump, but the maser
indicates a likely very young internal source.
Clump \centerclump to the east contains a NIR-red source unresolved by
HST, which is a bright MIR source (Fig.~\ref{overviewfig}), identified
as a MYSO by \citet{gc09}.  It also is associated with 22GHz maser
emission at the same velocity as the CO, but the CO emission is
extended in the north/south direction, and the maser position is
2.0\arcsec\  north of the NIR source \citep{imai}.  The mid-IR MYSO is
discussed as ``S9'' in \citet{walborn13}.

Clump~\eastclump contains a 22GHz H$_2$O maser \citep{imai} and a very
red NIR source unresolved by HST.  The NIR source is a bright MIR
\citet{gc09} MYSO.  Furthermore, this clump has the broadest linewidth
of any in our data, probably due to outflows from the embedded
MYSO(s).

Cluster P4 \citep{hyland92} to the north is unassociated with any CO
emission peak, so we suggest that it is also relatively evolved and
has largely cleared its birth material (although not completely, as
evidenced by the filament seen in absorption on the northwest side of
the stellar cluster).  \citet{walborn13} designate the brightest MIR
source in the cluster ``S8''.

% ==============================================================
\section{Analysis: Mass Distribution}
\label{mass}

% ==============================================================
\subsection{Molecular Column Density and Mass Derived from CO}
\label{ltemass}

Column density was calculated from $^{12}$CO and \co{13}{2-1} using
common assumptions \citep{rohlfs}, described in this section along with their
caveats.  If \co{12}{2-1} is very optically thick, the excitation
temperature T$_{ex}$ is only a function of the brightness temperature.
\[ T_{ex} =
{{11.1K}\over{\ln\left({{11.1}\over{I_{12}+0.19}}+1\right)}},\label{Tex}\] where
I$_{12}$ is the \co{12}{2-1} intensity in K.
If it is further assumed that $^{13}$CO has the same excitation
temperature as $^{12}$CO, the relative abundance is constant, and that
CO is thermalized (the level population is described by the Boltzmann
distribution evaluated at the excitation temperature), then the column
density per velocity bin can be calculated:

\begin{eqnarray*}
\tau_0^{13} &=& -\ln\left[1-{{T_B^{13}}\over{10.6}}\left\{ {1\over{e^{10.6/T_{ex}}-1}} - {1\over{e^{10.6/2.7}-1}} \right\}^{-1} \right] \\
N(^{13}{CO}) &=& 1.5\times 10^{14} { {T_{ex} e^{5.3/T_{ex}}\int \tau^{13}_v dv}\over{1-e^{-10.6/T_{ex}}}} 
\label{NCO}
\end{eqnarray*}

\begin{figure}[h!]
%\resizebox{3.6in}{!}{\includegraphics{30Dor_12CO_Tex.eps}}
%\resizebox{3.6in}{!}{\includegraphics{30Dor_13CO_Ntot.eps}}
\resizebox{3.in}{!}{\includegraphics{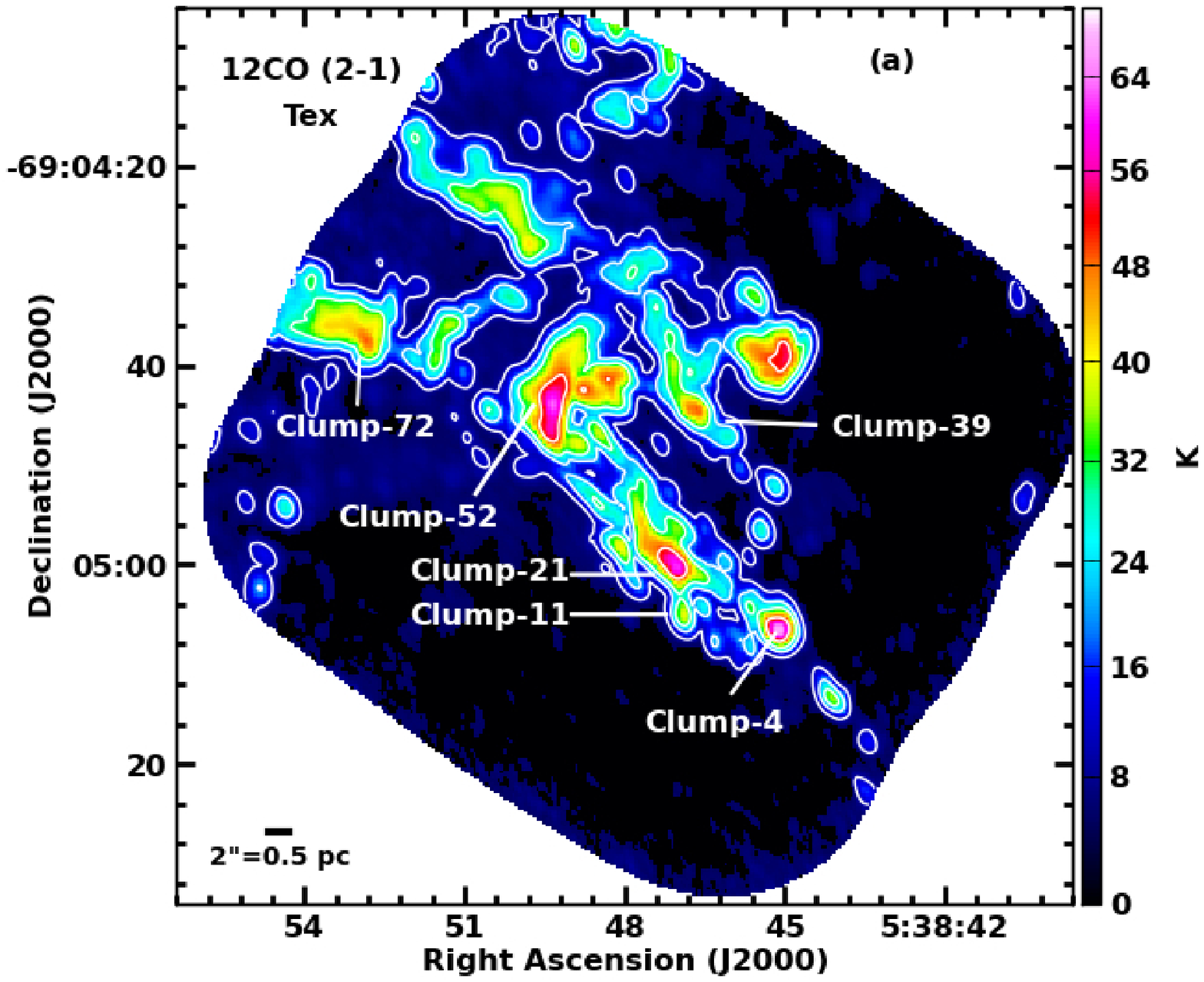}}
\resizebox{3.in}{!}{\includegraphics{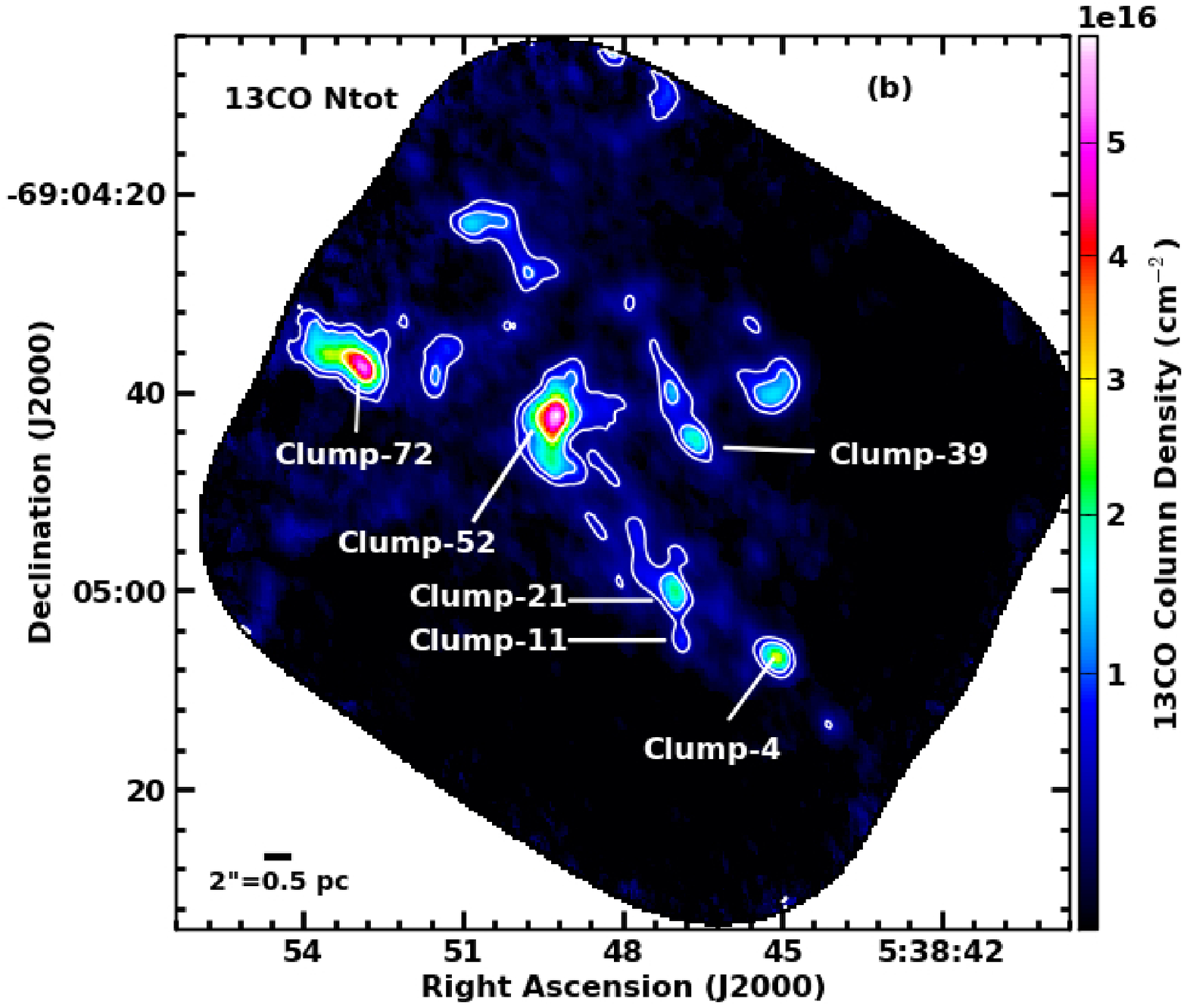}}
\caption{T$_{ex}$ calculated from \co{12}{2-1} brightness temperature, and $^{13}$CO column density calculated from that T$_{ex}$ and the \co{13}{2-1} intensity.\label{col}}
\end{figure}

Figure~\ref{col} shows the calculated excitation temperature and
column densities under these assumptions (the validity of which we
consider below).  The peak $^{13}$CO column density is
$\sim$5$\times$10$^{16}$cm$^{-2}$ in the central clump~\centerclump.  Numerous
clumps exceed a $^{13}$CO column density of 10$^{16}$cm$^{-2}$ (H$_2$
column density of 10$^{23}$cm$^{-2}$ with the abundances discussed
below).  Models of low-spatial-resolution CO emission
(\S\ref{introduction}) agree on a low filling factor of dense gas,
$\lesssim$15\% and as low as 1.5\% in the LVG models of
\citet{nikolic07}.  This can now be compared to our high-resolution
data.  The fraction of our map with
N($^{13}$CO)$\ge$5$\times$10$^{16}$cm$^{-2}$ is less than 1\%, quite
consistent with the low-resolution LVG parameters.

The assumptions of this ``molecular column density'' or ``LTE mass''
calculation have been considered by numerous authors in the past
\citep[e.g.][]{koppen80,heyer09}.  The higher abundance of $^{12}$CO
relative to $^{13}$CO means that its effective critical density
(including radiative trapping) is lower.  For densities
n(H$_2$)$\lesssim$1000$\,$cm$^{-3}$ (depending on column density),
$^{12}$CO will be largely thermalized, whereas $^{13}$CO remains
subthermally excited, lowering the calculated $^{13}$CO column density
\citep[these statements are true for 1-0 and 2-1 as one can verify
with RADEX; ][]{radex}.
However, when subthermally excited, the partition function
overestimates the population of higher rotational levels, and the
calculated total $^{13}$CO column density could be incorrectly high.
\citet{heyer09} illustrate this effect for CO$\;$1-0 in relatively
dark clouds, showing that the LTE method can underestimate the column
in low-density (n(H$_2)\simeq$500cm$^{-3}$) envelopes by a factor of
two and overestimate it for
10$^3\lesssim$n(H$_2$)$\lesssim$10$^4$cm$^{-3}$.  \citet{padoan00}
concluded from numerical simulations that LTE mass can underestimate
the true mass by a factor of a few, but the effect is largest at
$^{13}$CO column densities $\lesssim$10$^{14}$cm$^{-2}$ and H$_2$
volume densities $\lesssim$10$^{3}$cm$^{-3}$ As we will determine in
\S\ref{clumps}, most of the dense clumps in \dor\ have column
densities $\gtrsim$10$^{14}$cm$^{-2}$, and volume densities
10$^{4}<$n(H$_2$)$<$10$^{6}$cm$^{-3}$. We expect the analysis to be
robust, but that individual masses may have uncertainties of a factor
of two.

Observations of CO$\;$2-1 on large (100-1000pc) scales indicates that
the factor of two higher critical density makes it a moderately less
robust tracer of bulk molecular gas mass than CO$\;$1-0
\citep{koda12,sawada01,sakamoto97}. However this occurs because most
of a GMC has relatively {\em low} column and volume density -- the
clumps and dense warm molecular gas observed at much smaller scales
here are well-traced by CO$\;$2-1.  Analysis of CO$\;$2-1 by
\citet{koppen80} finds that the LTE analysis is appropriate for
n(H$_2$)$\gtrsim$10$^4$ and $^{12}$CO optical depth $\gtrsim$1,
i.e. brightness temperature ratio
T$_B$(\co{12}{2-1})/T$_B$(\co{13}{2-1}) $\gtrsim$20.  The latter
condition is violated in only a vanishingly small fraction of our
$^{13}$CO cube ($<$0.5\%), on the edges of clumps.  Since we primarily
compare clumps in 30Dor-10 with those calculated with the same method
in other clouds, we note these possible systematic errors but do not
expect them to affect our conclusions.

Another effect, of particular importance in regions of high radiation
intensity, is that of differential abundance, or isotopic
fractionation.  In diffuse gas (A$_V<<$ 1) lower $^{13}$CO column
density results in lower self-shielding and a reduced abundance
relative to $^{12}$CO \citep[e.g.][]{bally82}.  In dense PDRs and
clouds, this effect is counteracted by ion-molecule fractionation
$^{13}C^+ + ^{12}CO \leftrightarrow ^{12}C^+ + ^{13}CO + 36K$.
Selective photodissociation only dominates in optically thin diffuse
molecular gas, whereas for the conditions in 30Dor-10, $^{13}$CO is
chemically enhanced at the $A_V\lesssim$1 surface
%\citep{vdblack88,liszt98,rollig13}, 
and fractionation is minimal
deeper in the cloud where most of the observed photons are emitted
\citep{visser09,rollig13}.

Next, one must consider that $^{12}$CO becomes optically thick near
the surface of a PDR (within N(H$_2$)$\sim$10$^{20}$ of the transition from C$^+$ to C$^0$ to CO, typically at $A_V\sim$2 for the Milky
Way gas to dust ratio), whereas $^{13}$CO remains at most moderately
optically thick throughout a dense clump.  Even if the lower $J$
levels are thermalized, the kinetic temperature is quite high on the
surface of a PDR (several 100K).  One may question whether the
$^{12}$CO excitation temperature is thus representative of most of the
$^{13}$CO-emitting gas.  We estimate from PDR models that the kinetic
temperature at the $^{12}$CO photosphere is no more than $\sim$1.5
times the bulk of $^{13}$CO emitting gas \citep{teilholl,rollig13},
resulting in less than $\sim$10\% error in the calculated CO column
densities, unless the center of the clump is very cold $<$10K.

Converting the measured column densities to H$_2$ column densities
requires two abundance ratios: $^{12}$CO/$^{13}$CO, and
H$_2$/$^{12}$CO.  The $^{12}$CO/$^{13}$CO ratio is 70 in the solar
neighborhood; atomic isotopic ratios $^{12}$C/$^{13}$C and
$^{16}$O/$^{18}$O increase with decreasing metallicity in the outer
Milky Way \citep{milam05,wilsonrood}, but the molecular ratio is less
well constrained.  Previous measurements of $^{12}$CO/$^{13}$CO in
30Dor-10 range from 38 \citep{heikkila} to 50--100 \citep{nikolic07}.
The H$_2$/CO abundance depends on the cloud temperature, density, and
radiation field. Studies at parsec scales of outer Milky Way molecular
clouds (at similar 1/2-1/3 metallicity to the LMC) use H$_2$/$^{13}$CO
of 10$^6$ \citep{heyer01} to 3$\times$10$^6$ \citep{brand95}.  The
value should increase in photodissociation regions because of
selective photodissociation of CO relative to H$_2$ -- PDR models
suggest H$_2$/$^{13}$CO $\gtrsim$10$^7$ even at solar metallicity
\citep{vd87}, but our analysis shows below that these clumps are
better described as dense molecular condensations and not PDRs.  We
will use H$_2$/$^{13}$CO=5$\times$10$^6$, which as shown in
section~\ref{massmass} yields good agreement between several methods
of calculating H$_2$ mass.

%=================================================================
\subsection{H$_2$ Mass from Dust Continuum Emission}
\label{continuum}

Once the contribution from thermal bremsstrahlung has been removed,
the remaining millimeterwave continuum emission arises from dust, which
provides an independent measure of the gas column density and mass.
The intensity I$_\nu$ of dust emission is given by:
\[ I_\nu = \Sigma_{gas}\kappa_\nu x_d B_\nu(T_{dust}), \] where
$\Sigma_{gas}$ is the gas mass surface density, $\kappa_\nu$ the dust
emissivity per unit dust mass, and $x_d$ the dust-to-gas ratio by mass.
We adopt $\kappa_d$=0.62$\pm$0.1$\;$cm$^2\;$g$^{-1}$, midway between
0.72$\;$cm$^2\;$g$^{-1}$ used by \citet{bot07} in the SMC, and
0.52$\;$cm$^2\;$g$^{-1}$ used by \citet{galliano11} in the LMC.  The
value is 2 times higher than \citet{dl84}, so our dust masses could be
low.  Scaling the gas-to-dust ratio by the LMC metallicity of 1/2
solar \citep{rolleston03,pagel03} yields $x_d$=1/339
\citep{galliano11}. It is not well known the extent to which the
gas-to-dust ratio is uniform at such small scales -- \citet{lebou08}
found significant variations of the Fe/H abundance on parsec scales in
30~Doradus which might be due to dust destruction.  Proceeding with
this caveat, the gas surface density in cm$^2\;$g$^{-1}$ is
\[ \Sigma_{gas} = {{I_{\nu,cont}}\over{\rm Jy\;bm^{-1}}} 230 \left( e^{h(\nu,cont)/kT_{dust}}-1 \right). \]
If we assume that the CO excitation temperature calculated above from
the \co{12}{2-1} brightness temperature equals the dust temperature,
we can easily make a column density map.  
Note that the dust temperature derived from {\it Herschel} data by
Gordon et al. (2014) in prep. is 55K in the single resolution element
containing 30Dor-10, so at least in spatial average, the dust
temperature is similar to the gas excitation temperature shown in
Figure~\ref{col}.
Furthermore, since h($\nu$,cont)/k=10.8$\;$K is
very close to h($\nu$,\co{12}{2-1})/k = 11.1$\;$K, we can replace
the exponential term with 11.1/T$_{B,12CO}$ to an accuracy of better
than 98\%, and $\Sigma_{gas} = 2500\;$cm$^3\;$g$^{-1}$I$_{\nu,cont}$/T$_{B,12CO}$.

The total mass in the map derived from dust emission is
6$\pm$1$\times$10$^4\;$M$_\odot$, consistent with
3.6$\times$10$^4\;$M$_\odot$ and 3.8$\times$10$^5\;$M$_\odot$
calculated by \citet{johansson98} from \co{12}{1-0} luminosity and the
virial theorem, respectively (their measurements apply to the entire
30Dor-10 cloud, and they don't use the optically thin isotopologue to
calculate molecular mass, so the estimates should only be expected to
agree to a factor of several).  Masses of individual clumps in our
data are analyzed next.

% ==============================================================
\section{Analysis: Dense Clumps in the 30Dor-10 Molecular Cloud}
\label{clumps}

% SUB2
Emission from complex molecular clouds is often analyzed by
decomposition into clouds or clumps, which by construction isolates a
particular size scale a few times the spatial beam.  In this section
we describe the decomposition, and then how 30Dor-10 clumps compare to
previously studied molecular clouds in size-linewidth-surface density
parameter space.  
%Low-J CO emission such as used here is sensitive to structures with volume densities similar to its critical density, $\sim$10$^3$cm$^{-3}$, so 
Higher resolution will likely break these clumps up into smaller
substructures, but we can gain insight by comparing physical
parameters derived from these data with clumps in other clouds
decomposed at similar physical resolution.
%/SUB2

\subsection{Cloud Decomposition}
\label{clumpdecomp}

% SUB2
Different decomposition algorithms have been explored, for example {\tt gaussclumps} \citep{stutzki}, in which Gaussians are fit to the brightest peaks, subtracted, and the process iterated.  Such fitting allows the separation of blended structures, but assumes a functional form for the intensity (Gaussians).  In extragalactic studies, methods which do not presuppose a functional form have been more common, in particular
% /SUB2
%Two programs commonly used to decompose emission cubes into
%substructures or clumps are 
{\tt clumpfind} \citep{williams} and {\tt
  cprops} \citep{cprops_pasp}.  Both methods select local maxima in
the data cube and then assign pixels to those local maxima.  The {\tt
  clumpfind} program assigns all pixels with flux above some specified floor
to the nearest local maximum, whereas {\tt cprops} only assigns pixels
within an isosurface which contains a single local maximum.  In
crowded regions with complex emission, {\tt clumpfind} tends to
produce a ``patchwork quilt'' assignment cube dividing up the low
level emission, whereas {\tt cprops} only assigns the peaks, leaving
fairly bright emission unassigned.

We use a hybrid approach in which local maxima are chosen using the
{\tt cprops} criteria, based on their contrast above shared emission,
and signal-to-noise. Then emission between local maxima is assigned,
down to a flux level containing several local maxima, but not all the
way down to the same noise floor across the entire region.  All of the
trends discussed below are calculated for a range of decomposition
parameters, both with and without the apportionment of emission using
the hybrid approach.  Although some of the fainter individual clumps
properties change, all of the trends below are robust to the
decomposition method.  

Clump properties are measured from the intensity-weighted moments in space and
velocity of the pixels assigned to each clump. These moments are deconvolved by
the synthesized beam and velocity channel width, respectively.  The spatial
moment must be converted to a radius, which depends in the intensity profile.
The radius of a Gaussian beam is often defined as r=2$\sigma_x$, the 1/e$^2$
point, or an area containing 68\% of the flux.  We adopt the commonly used
convention for a radial density profile $\rho\propto$r$^{-1}$ which yields
r=1.91$\sigma_x$ \citep{srby,bolatto_araa}.  Clump properties are listed in
Table~\ref{properties}.  Radii range from unresolved up to $\sim$0.6pc, with a
median for nominally resolved clumps of 0.3pc.
% SUB2
The minimum fittable clump size depends on the signal-to-noise as described in
\citet{condon97}.  For the faint clumps in our sample (0.75K = 5$\sigma$ for
\co{13}{2-1}), the uncertainty in radius if a Gaussian were fitted to the clump
is 0.35\arcsec = 0.09$\;$pc, decreasing linearly with increasing brighness.  We
do not use any deconvolved radii less than 0.1$\;$pc.
% /SUB2
Velocity dispersions $\sigma_v$ range between 0.5 and
3$\;$\kms, with a median value of 1.1$\;$\kms (FWHM 2.6$\;$\kms).
Approximate volume densities M/(4/3$\pi$r$^3$) range from 10$^3$ to
10$^5\;$cm$^{-3}$, with a median value of 10$^4\;$cm$^{-3}$.

\begin{deluxetable}{lllllllllllllllllll}
\rotate
\tabletypesize{\scriptsize}
\tablecolumns{18}
\tablewidth{0pc}
\renewcommand{\tabcolsep}{1ex}
\tablecaption{Clump Properties\label{properties}}
\tablehead{ & \multicolumn{2}{c}{position} & velocity & radius & $\delta$r & $\sigma_v$ & $\delta\sigma_v$ & $^{13}$CO$_{pk}$ & F($^{13}$CO) & $\delta$F$_{13}$ & $^{12}$CO$_{pk}$ & F($^{12}$CO) & $\delta$F$_{12}$ & M$_{mol}$ & $\delta$M$_{mol}$ & M$_{vir}$ & $\delta$M$_{vir}$ & \\
& RA (J2000) & Dec (J2000) & \kms & \multicolumn{2}{c}{pc} & \multicolumn{2}{c}{\kms} & K & \multicolumn{2}{c}{K$\;$\kms$\;$pc$^2$} & K & \multicolumn{2}{c}{K$\;$\kms$\;$pc$^2$} & \multicolumn{2}{c}{M$_\odot$} & \multicolumn{2}{c}{M$_\odot$} & }
\startdata
  1& 5:38:40.53& -69:04:53.5&  262.4&  0.11& 0.03&  0.89& 0.21&   2.5&  1.5&  0.3& 10.6&   5.6&   0.6&      84&    19&    90&    35& \\
  2& 5:38:40.77& -69:04:32.7&  263.8&  0.11& 0.05&  0.85& 0.59&   1.7&  0.6&  0.6&  8.6&   1.3&   0.3&      34&    37&    83&    67& \\
  3& 5:38:44.08& -69:05:12.9&  250.7&  0.18& 0.04&  0.85& 0.17&   5.5&  4.6&  0.1& 30.9&  30.3&   0.1&     307&     7&   136&    39& \\
  4& 5:38:45.12& -69:05:06.6&  248.3&  0.29& 0.04&  1.36& 0.08&  13.9& 22.8&  0.1& 61.8& 166.7&   0.1&    2256&    13&   559&    91& pillar, cluster K1 \\
  5& 5:38:45.65& -69:05:04.9&  251.1&  0.22& 0.04&  0.42& 0.08&   3.2&  3.8&  0.1& 27.3&  33.5&   0.1&     260&     8&    40&    11& \\
  6& 5:38:45.68& -69:05:03.3&  247.9&  0.11& 0.03&  0.68& 0.13&   2.1&  0.8&  0.4& 20.1&  11.5&   0.4&      51&    25&    52&    17& \\
  7& 5:38:45.65& -69:05:08.5&  251.1&  0.11& 0.02&  0.72& 0.13&   3.8&  1.9&  0.2& 19.0&  12.4&   0.2&     159&    17&    59&    14& \\
  8& 5:38:45.81& -69:05:08.7&  246.3&  0.13& 0.06&  0.93& 0.22&   1.4&  1.1&  0.2& 15.1&  10.7&   0.2&      85&    11&   116&    57& \\
  9& 5:38:46.04& -69:05:06.6&  245.3&  0.11& 0.01&  0.57& 0.08&   1.7&  0.4&  0.2& 14.1&   3.8&   0.2&      29&    14&    37&     7& \\
 10& 5:38:46.32& -69:05:07.7&  245.6&  0.11& 0.04&  0.92& 0.08&   2.2&  1.4&  0.2& 16.3&  10.6&   0.1&     107&    13&    97&    33& \\
 11& 5:38:46.88& -69:05:04.9&  248.2&  0.16& 0.02&  1.25& 0.12&   6.6&  7.5&  0.1& 33.8&  52.5&   0.1&     540&     8&   260&    40& pillar, NIR YSO \\
 12& 5:38:46.58& -69:05:04.1&  245.6&  0.19& 0.04&  0.59& 0.23&   3.2&  1.3&  0.2& 20.2&  10.3&   0.2&      90&    13&    68&    30& \\
 13& 5:38:46.33& -69:05:02.3&  248.1&  0.18& 0.04&  1.00& 0.13&   4.9&  3.0&  0.2& 32.8&  25.1&   0.1&     239&    13&   185&    49& \\
 14& 5:38:46.17& -69:05:01.7&  253.4&  0.11& 0.04&  0.75& 0.23&   3.8&  1.9&  0.2& 18.9&  14.2&   0.2&     162&    18&    63&    29& \\
 15& 5:38:44.69& -69:04:57.9&  234.5&  0.11& 0.04&  0.43& 0.13&   0.9&  0.2&  0.7&  3.3&   0.6&   0.3&      19&    61&    21&     9& \\
 16& 5:38:44.52& -69:04:57.7&  240.0&  0.11& 0.07&  1.59& 0.30&   0.8&  0.4&  0.9&  4.2&   1.5&   0.2&      34&    72&   300&   202& \\
 17& 5:38:44.23& -69:04:55.6&  251.9&  0.11& 0.05&  0.81& 0.21&   0.7&  0.2&  1.0&  2.1&   0.5&   0.4&      14&    57&    74&    41& \\
 18& 5:38:45.45& -69:04:56.3&  242.5&  0.15& 0.04&  0.65& 0.06&   5.0&  2.7&  0.1& 22.5&  14.7&   0.1&     194&    10&    65&    17& \\
 19& 5:38:45.21& -69:04:52.0&  243.1&  0.27& 0.06&  0.55& 0.04&   4.3&  2.5&  0.1& 21.4&  13.8&   0.2&     169&     9&    86&    21& \\
 20& 5:38:45.80& -69:04:50.5&  243.8&  0.22& 0.06&  0.62& 0.08&   2.0&  1.2&  0.2& 12.8&   5.6&   0.2&      61&    11&    88&    26& \\
 21& 5:38:47.05& -69:05:00.3&  250.1&  0.41& 0.08&  0.93& 0.06&  16.5& 20.4&  0.1& 55.7& 128.8&   0.1&    2000&    12&   375&    73& pillar, NIR YSO \\
 22& 5:38:47.60& -69:04:57.9&  251.5&  0.25& 0.08&  0.85& 0.13&   7.3&  5.8&  0.1& 45.8&  41.4&   0.1&     519&    12&   185&    69& \\
 23& 5:38:47.09& -69:04:56.7&  251.2&  0.11& 0.06&  1.04& 0.04&   5.1&  3.9&  0.1& 32.3&  33.8&   0.1&     288&     9&   123&    65& \\
 24& 5:38:47.53& -69:05:00.4&  254.1&  0.11& 0.05&  1.27& 0.30&   1.3&  0.9&  0.2& 12.7&   8.3&   0.2&      76&    13&   185&    95& \\
 25& 5:38:47.76& -69:05:02.0&  243.7&  0.11& 0.05&  0.85& 0.04&   2.7&  1.5&  0.2& 19.6&  11.2&   0.1&     106&    13&    82&    39& \\
 26& 5:38:48.16& -69:04:58.1&  243.3&  0.28& 0.04&  0.79& 0.04&   3.8&  4.0&  0.1& 35.3&  45.1&   0.1&     315&    10&   180&    29& \\
 27& 5:38:46.47& -69:04:56.6&  247.5&  0.11& 0.03&  1.27& 0.25&   1.9&  1.4&  0.3& 12.2&   6.8&   0.3&      70&    15&   185&    69& \\
 28& 5:38:46.90& -69:04:56.4&  243.5&  0.10& 0.04&  0.88& 0.04&   4.7&  3.1&  0.1& 20.6&  17.7&   0.1&     200&     8&    89&    30& \\
 29& 5:38:48.36& -69:05:00.6&  254.9&  0.14& 0.08&  2.35& 0.85&   0.7&  0.5&  1.1&  6.0&   2.9&   0.3&      41&    98&   810&   525& \\
 30& 5:38:47.15& -69:04:53.1&  241.6&  0.11& 0.04&  0.51& 0.04&   2.1&  0.8&  0.4& 12.6&   5.5&   0.3&      51&    25&    29&     9& \\
 31& 5:38:47.65& -69:04:54.6&  248.8&  0.17& 0.08&  1.27& 0.30&   2.6&  1.8&  0.1& 24.5&  18.2&   0.1&     131&    10&   280&   146& \\
 32& 5:38:47.75& -69:04:54.0&  252.1&  0.33& 0.14&  1.10& 0.30&   6.5&  5.3&  0.2& 37.8&  39.1&   0.3&     418&    13&   418&   214& \\
 33& 5:38:48.49& -69:04:53.8&  245.8&  0.36& 0.16&  1.10& 0.21&   5.2&  7.9&  0.1& 25.9&  65.7&   0.1&     610&     8&   457&   222& \\
 34& 5:38:47.73& -69:04:50.9&  250.2&  0.21& 0.07&  1.15& 0.15&   4.1&  3.6&  0.1& 27.1&  27.2&   0.1&     274&    10&   290&   108& \\
 35& 5:38:48.35& -69:04:47.8&  251.0&  0.25& 0.14&  1.10& 0.21&   6.5&  8.7&  0.1& 32.6&  51.4&   0.1&     588&     6&   315&   182& \\
 36& 5:38:47.28& -69:04:47.5&  241.0&  0.11& 0.03&  0.85& 0.42&   1.1&  0.3&  0.5&  8.8&   3.1&   0.2&      27&    48&    82&    46& \\
 37& 5:38:46.38& -69:04:48.7&  241.1&  0.19& 0.06&  1.21& 0.08&   1.2&  1.1&  0.2&  7.7&   5.2&   0.4&      61&    14&   286&    91& \\
 38& 5:38:46.18& -69:04:47.2&  252.5&  0.11& 0.04&  1.27& 0.21&   4.8&  4.0&  0.3& 25.2&  31.9&   0.1&     309&    22&   185&    68& \\
 39& 5:38:46.71& -69:04:45.0&  254.5&  0.28& 0.09&  1.21& 0.06&  12.7& 18.0&  0.3& 43.2& 118.1&   0.1&    1652&    24&   432&   134& H$_2$O maser \\
 40& 5:38:47.13& -69:04:39.5&  253.3&  0.32& 0.10&  1.78& 0.17&   7.4& 14.4&  0.1& 32.2& 114.7&   0.1&    1168&     7&  1048&   360& \\
 41& 5:38:46.49& -69:04:40.2&  252.4&  0.26& 0.06&  0.98& 0.11&   2.4&  2.4&  0.2& 22.9&  31.0&   0.1&     152&    12&   259&    69& \\
 42& 5:38:46.76& -69:04:41.2&  248.0&  0.33& 0.09&  1.70& 0.42&   2.5&  4.2&  0.1& 25.3&  52.7&   0.1&     303&     6&  1003&   373& \\
 43& 5:38:47.17& -69:04:42.1&  250.6&  0.11& 0.04&  1.91& 0.42&   3.0&  3.6&  0.1& 27.9&  39.2&   0.1&     254&     9&   417&   180& \\
 44& 5:38:44.95& -69:04:38.9&  251.5&  0.38& 0.08&  1.10& 0.08&  10.4& 16.3&  0.1& 48.6&  97.0&   0.1&    1509&    11&   481&   107& \\
 45& 5:38:45.39& -69:04:40.6&  250.0&  0.22& 0.06&  1.06& 0.30&   8.4&  7.3&  0.2& 42.6&  63.1&   0.1&     646&    17&   259&   103& \\
 46& 5:38:45.43& -69:04:39.7&  252.4&  0.35& 0.10&  1.06& 0.30&   8.2&  8.8&  0.1& 47.1&  57.8&   0.1&     795&    10&   415&   166& \\
 47& 5:38:45.85& -69:04:38.3&  253.3&  0.13& 0.07&  1.02& 0.17&   5.6&  2.4&  0.2& 35.5&  18.5&   0.2&     153&    11&   144&    84& \\
 48& 5:38:45.60& -69:04:33.3&  250.6&  0.35& 0.07&  0.98& 0.13&   6.9&  7.5&  0.1& 29.5&  62.0&   0.1&     515&     8&   345&    83& \\
 49& 5:38:47.78& -69:04:41.4&  257.7&  0.24& 0.06&  1.49& 0.17&   4.5&  3.2&  0.1& 30.6&  26.0&   0.1&     265&    11&   546&   156& \\
 50& 5:38:48.23& -69:04:41.6&  255.1&  0.39& 0.08&  1.15& 0.13&   7.3&  8.2&  0.1& 44.8&  67.3&   0.2&     655&    11&   529&   122& \\
 51& 5:38:48.74& -69:04:42.6&  253.4&  0.40& 0.07&  1.15& 0.17&   6.1&  9.0&  0.1& 45.0&  71.2&   0.3&     709&    10&   543&   129& \\
 52& 5:38:49.35& -69:04:42.4&  250.5&  0.53& 0.09&  1.70& 0.21&  23.9& 81.5&  0.1& 57.8& 371.1&   0.1&    8528&     8&  1585&   340& {\em Spitzer} MYSO, H$_2$O maser \\
 53& 5:38:49.41& -69:04:46.8&  249.7&  0.39& 0.10&  1.70& 0.21&  11.6& 21.6&  0.1& 51.9& 136.0&   0.1&    1934&     8&  1165&   345& \\
 54& 5:38:49.00& -69:04:38.6&  249.3&  0.41& 0.10&  1.91& 0.08&   8.8& 13.7&  0.1& 36.7& 118.0&   0.1&    1102&     6&  1539&   387& \\
 55& 5:38:48.95& -69:04:47.9&  253.1&  0.36& 0.10&  1.59& 0.08&   7.3&  9.8&  0.1& 34.2&  61.3&   0.2&     654&     7&   936&   268& \\
 56& 5:38:51.13& -69:04:51.1&  250.9&  0.11& 0.08&  1.49& 0.64&   0.9&  0.3&  1.6&  4.4&   2.3&   0.3&      36&   167&   252&   219& \\
 57& 5:38:50.14& -69:04:45.7&  245.3&  0.11& 0.03&  0.89& 0.08&   1.9&  1.3&  0.4& 11.8&   4.5&   0.6&      56&    17&    90&    29& \\
 58& 5:38:50.49& -69:04:44.5&  253.8&  0.11& 0.03&  0.54& 0.06&   3.4&  1.6&  0.2& 25.3&  11.4&   0.4&      74&     9&    32&    10& \\
 59& 5:38:47.43& -69:04:34.8&  253.6&  0.39& 0.08&  1.91& 0.17&   3.2&  7.3&  0.1& 25.2&  64.8&   0.1&     472&     6&  1468&   327& \\
 60& 5:38:48.13& -69:04:31.2&  249.1&  0.30& 0.15&  2.76& 0.85&   1.7&  3.5&  0.1& 19.5&  72.6&   0.1&     248&     5&  2363&  1421& \\
 61& 5:38:47.86& -69:04:30.8&  243.6&  0.30& 0.04&  2.76& 0.85&   2.0&  4.6&  0.1& 25.8&  88.8&   0.1&     341&     5&  2385&   813& \\
 62& 5:38:46.63& -69:04:30.5&  239.1&  0.08& 0.07&  0.64& 0.42&   0.6&  0.3&  0.8&  8.6&   2.0&   0.3&      15&    42&    46&    43& \\
 63& 5:38:47.48& -69:04:29.6&  239.3&  0.24& 0.04&  1.91& 0.21&   2.5&  4.5&  0.1& 22.5&  46.3&   0.1&     229&     4&   926&   192& \\
 64& 5:38:48.59& -69:04:30.9&  259.6&  0.11& 0.04&  0.98& 0.13&   1.2&  0.9&  0.2& 11.9&   8.0&   0.2&      55&    15&   109&    40& \\
 65& 5:38:52.23& -69:04:50.8&  255.4&  0.33& 0.11&  2.59& 0.17&   1.1&  1.4&  0.1&  3.7&  10.5&   0.2&     222&    15&  2306&   813& \\
 66& 5:38:54.85& -69:05:02.2&  262.2&  0.23& 0.09&  0.66& 0.06&   2.1&  1.6&  0.6& 15.2&   4.6&   0.2&      55&    22&   104&    41& \\
 67& 5:38:54.71& -69:04:59.4&  260.2&  0.35& 0.07&  1.47& 0.06&   1.8&  1.2&  2.0&  9.0&   6.1&   0.3&      95&   154&   789&   168& \\
 68& 5:38:54.29& -69:04:54.2&  260.2&  0.11& 0.05&  0.65& 0.04&   5.6&  3.4&  0.1& 22.6&  16.8&   0.1&     249&    10&    48&    24& \\
 69& 5:38:54.85& -69:04:45.8&  256.4&  0.16& 0.10&  0.85& 0.21&   1.6&  0.8&  0.6&  7.8&   3.4&   0.2&      97&    72&   119&    77& \\
 70& 5:38:53.88& -69:04:42.4&  263.9&  0.16& 0.15&  0.55& 0.13&   1.2&  0.7&  0.2&  8.4&   4.7&   0.3&      68&    15&    50&    48& \\
 71& 5:38:53.42& -69:04:40.5&  264.5&  0.11& 0.10&  1.15& 0.17&   1.2&  0.7&  0.2&  7.4&   5.1&   0.1&      70&    15&   150&   143& \\
 72& 5:38:52.94& -69:04:37.4&  254.1&  0.49& 0.12&  2.89& 0.08&  15.3& 86.1&  0.1& 43.4& 409.9&   0.0&    7375&     4&  4262&  1061& {\em Spitzer} MYSO, H$_2$O maser \\
 73& 5:38:53.72& -69:04:35.6&  256.3&  0.65& 0.13&  1.94& 0.08&  12.6& 48.9&  0.1& 37.0& 227.1&   0.1&    3564&     4&  2525&   517& \\
 74& 5:38:52.62& -69:04:34.3&  260.1&  0.61& 0.18&  2.55& 0.85&   2.7&  5.4&  0.2& 24.6&  43.8&   0.1&     326&    11&  4123&  1816& \\
 75& 5:38:53.80& -69:04:32.5&  254.7&  0.07& 0.04&  3.18& 0.85&   4.6&  5.0&  0.4& 28.0&  46.5&   0.2&     273&    20&  1160&   507& \\
 76& 5:38:51.85& -69:04:39.5&  257.3&  0.11& 0.05&  0.62& 0.08&   1.7&  0.6&  0.2& 20.4&  10.8&   0.2&      47&    17&    43&    19& \\
 77& 5:38:51.85& -69:04:38.3&  253.7&  0.11& 0.05&  0.85& 0.34&   2.7&  1.4&  0.2& 19.7&  11.8&   0.2&      92&    13&    82&    49& \\
 78& 5:38:51.55& -69:04:38.9&  250.1&  0.27& 0.06&  1.19& 0.17&   7.4&  8.1&  0.1& 30.3&  52.1&   0.1&     675&    11&   395&   110& \\
 79& 5:38:51.38& -69:04:36.6&  253.1&  0.47& 0.10&  0.89& 0.04&   7.8& 12.3&  0.1& 31.6&  73.7&   0.1&    1047&    12&   392&    83& \\
 80& 5:38:50.82& -69:04:34.7&  257.9&  0.37& 0.08&  1.17& 0.08&   2.4&  3.3&  0.2& 16.6&  29.5&   0.1&     224&    10&   517&   117& \\
 81& 5:38:50.08& -69:04:33.4&  259.5&  0.34& 0.07&  1.74& 0.06&   2.7&  6.0&  0.1& 24.4&  55.9&   0.1&     323&     4&  1080&   224& \\
 82& 5:38:49.34& -69:04:27.8&  246.1&  0.11& 0.04&  2.63& 0.30&   4.8&  5.9&  0.1& 28.9&  55.3&   0.1&     415&     8&   792&   282& \\
 83& 5:38:49.66& -69:04:25.8&  243.8&  0.11& 0.05&  2.89& 0.30&   1.8&  0.8&  0.2& 21.1&  10.8&   0.1&      56&    11&   953&   404& \\
 84& 5:38:48.88& -69:04:26.5&  243.3&  0.11& 0.10&  1.40& 0.08&   1.1&  0.9&  0.6& 15.4&  18.9&   0.3&      53&    36&   224&   208& \\
 85& 5:38:49.91& -69:04:27.2&  248.1&  0.44& 0.11&  1.26& 0.04&   8.1& 17.5&  0.1& 36.1& 132.7&   0.1&    1391&     7&   723&   191& \\
 86& 5:38:49.43& -69:04:23.2&  247.7&  0.35& 0.10&  1.36& 0.13&   1.4&  1.5&  0.2& 14.8&  26.4&   0.2&      98&    15&   678&   196& \\
 87& 5:38:50.26& -69:04:22.5&  243.7&  0.28& 0.11&  2.25& 0.21&   5.7& 10.9&  0.2& 34.1& 104.3&   0.1&     907&    18&  1469&   589& \\
 88& 5:38:50.69& -69:04:22.5&  245.0&  0.11& 0.09&  2.32& 0.21&   6.0&  6.9&  0.1& 32.0&  64.0&   0.1&     577&    10&   617&   496& \\
 89& 5:38:50.93& -69:04:23.7&  247.9&  0.28& 0.11&  1.70& 0.17&   7.8& 12.4&  0.1& 31.4&  93.6&   0.1&    1041&    10&   831&   343& \\
 90& 5:38:50.61& -69:04:20.2&  247.2&  0.14& 0.09&  2.87& 0.21&   2.4&  1.5&  0.1& 22.6&  22.1&   0.2&     106&    10&  1178&   806& \\
 91& 5:38:50.81& -69:04:19.1&  249.0&  0.11& 0.05&  1.87& 0.13&   2.3&  1.4&  0.2& 22.5&  17.6&   0.1&     106&    12&   399&   174& \\
 92& 5:38:50.95& -69:04:18.7&  245.3&  0.11& 0.05&  0.55& 0.13&   1.5&  0.6&  0.2& 21.3&   9.5&   0.2&      44&    16&    34&    16& \\
 93& 5:38:51.60& -69:04:21.6&  245.6&  0.11& 0.03&  1.23& 0.13&   2.0&  1.4&  0.1& 24.5&  26.8&   0.2&     107&    10&   173&    55& \\
 94& 5:38:51.89& -69:04:16.9&  246.8&  0.17& 0.06&  1.70& 0.06&   3.4&  4.0&  0.4& 25.4&  27.7&   0.2&     154&    14&   509&   181& \\
 95& 5:38:51.62& -69:04:12.1&  247.3&  0.11& 0.07&  0.79& 0.06&   5.6&  2.3&  0.2& 22.0&   6.9&   0.2&     150&    14&    72&    48& \\
 96& 5:38:49.76& -69:04:17.0&  240.9&  0.11& 0.10&  1.06& 0.21&   1.2&  0.5&  0.8& 10.8&   4.5&   0.2&      23&    38&   128&   116& \\
 97& 5:38:48.92& -69:04:17.6&  240.9&  0.11& 0.07&  2.12& 0.25&   0.9&  0.8&  0.5& 13.5&  16.1&   0.2&      48&    31&   515&   341& \\
 98& 5:38:48.17& -69:04:14.5&  248.7&  0.52& 0.12&  1.08& 0.06&   4.5&  6.6&  0.2& 23.7&  34.0&   0.1&     364&    10&   628&   147& near cluster P4\\
 99& 5:38:47.28& -69:04:11.7&  248.9&  0.32& 0.11&  0.68& 0.11&   7.9&  7.4&  0.1& 24.7&  28.8&   0.1&     509&    10&   152&    59& \\
100& 5:38:47.21& -69:04:11.5&  251.6&  0.11& 0.04&  0.64& 0.13&   2.2&  0.9&  0.7& 14.9&   4.0&   0.8&      41&    30&    46&    18& \\
101& 5:38:48.92& -69:04:08.0&  251.2&  0.11& 0.04&  0.76& 0.04&   4.3&  2.3&  0.2& 32.0&  16.8&   0.2&     133&    11&    66&    21& \\
102& 5:38:47.35& -69:04:10.1&  241.6&  0.11& 0.04&  0.56& 0.00&   1.8&  0.7&  1.0&  9.5&   1.7&   0.2&      44&    68&    36&    11& \\
103& 5:38:47.76& -69:04:13.7&  253.0&  0.11& 0.03&  0.76& 0.04&   1.8&  0.8&  0.4& 18.5&   8.8&   0.4&      41&    22&    66&    19& \\
\enddata
\end{deluxetable}

%=================================================================
\clearpage
\subsection{Scaling Relations}
\label{size-linewidth}

Observations of molecular clouds on $\gtrsim$10pc scales follow a
power-law scaling relation between the size $r$ and velocity
dispersion $\sigma_v\propto r^\beta$.  The index $\beta$ was 0.38 in the
original measurement \citep{larson}, and subsequently measured between
1/3 and 1/2 with high scatter.  The relation can be obtained from
incompressible turbulence ($\beta$=1/3), compressible shock dominated
turbulence ($\beta$=1/2), quasi-static gravitational equilibrium
($\beta$=1/2), or free-fall collapse ($\beta$=1/2) 
\citep[e.g.][]{turbreference}.  In this section
we analyze the 30Dor-10 clump distribution as a function of $r$,
$\sigma_v$, and surface density $\Sigma$.  
% SUB2 
These scaling relations are clearest when combining multiple datasets
at different spatial resolutions and using tracers sensitive to
different critical densities; our approach is to compare the clumps in
30Dor-10 to the trends derived from such multi-dataset analyses, and
to directly compare clumps in our data to clumps observed with similar
tracers and physical resolution in the Milky Way.
% /SUB2

\begin{figure}[h!]
%\resizebox{4in}{!}{\includegraphics{sizeline.eps}}
\resizebox{4in}{!}{\includegraphics{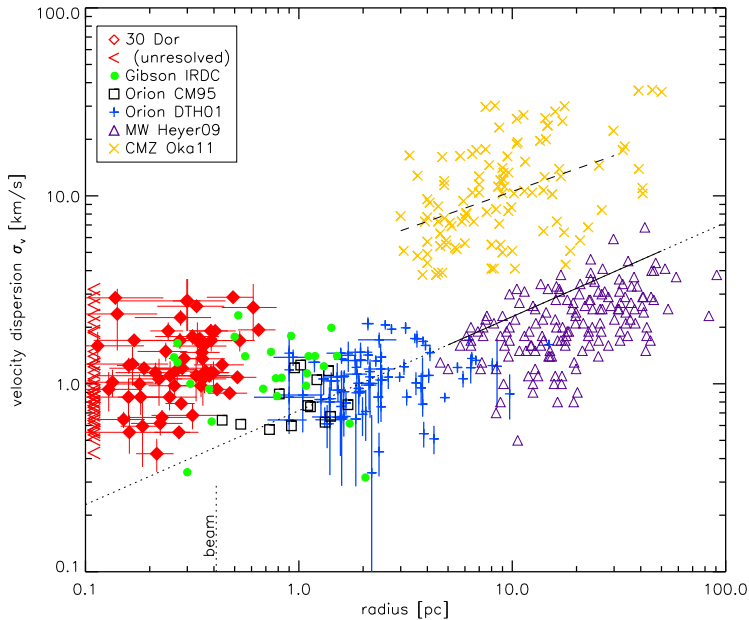}}
\caption{\label{sizeline} Size-linewidth relation for $^{13}$CO clumps in
  30Dor-10 compared to other locations.  Clumps are plotted as diamonds, or as
  left-facing triangles for those clumps that were not resolved in our
  observations (radius upper limits -- see text for discussion of deconvolving
  clump sizes).  Massive clumps in infrared-dark clouds are shown from Gibson et al. 2009 (green circles, ``Gibson IRDC''). Massive clumps in Orion are shown from Caselli \& Myers 1995
  (squares, ``Orion CM95'', \co{13}{1-0}), and also measured from the
  \co{12}{1-0} cube of \citet{dht01} using the same cloud decomposition method
  as we applied to 30Dor-10 (blue plus symbols, ``Orion DTH01'').  Milky Way clouds
  are taken from Heyer et al. 2009 (purple triangles, ``MW Heyer09'', $^{12}$CO and
  \co{13}{1-0}).  Those data follow the plotted relation
  $\sigma_v$=0.72R$^{0.5}$.  Also shown are \co{12}{1-0} clouds in the central
  molecular zone of the Milky Way from \citet{oka01} (orange crosses, ``CMZ Oka11''),
  and those authors' fitted relation as a dashed line.  See text for
  discussion.}
\end{figure}

The size-linewidth relation for $^{13}$CO clumps in 30Dor-10 is
compared to other locations in Figure~\ref{sizeline}.  Milky Way
clouds measured using $^{12}$CO and \co{13}{1-0} are taken from Heyer
et al. 2009, who revisited the seminal study of \citet[][hereafter
SRBY]{srby}.  The plotted relation $\sigma_v$=0.72R$^{0.5}$ is the fit
to the data reported by SRBY.  At somewhat smaller spatial scales, we
show structures in the Orion molecular cloud that we obtained using
the same cloud decomposition as in 30Dor-10, applied to the
\co{12}{1-0} data of \citet{dht01}.  Both sets of Orion clouds are
consistent with the SRBY Milky Way $\sigma_v$-$R$ relation.  Massive
clumps in Orion are also shown, from \citet[][\co{13}{1-0}]{cm95}. The
data in their table were converted from velocity FWHM to dispersion,
and half-width-half-max radii 1.18$\times\sigma_x$ to the radius
convention used here of 1.91$\times\sigma_x$.  Finally, clouds in the
central molecular zone (CMZ) of the Milky Way, measusing \co{12}{1-0}
by \citet{oka01} are plotted.  Clumps in 30Dor-10, and clouds in the
CMZ, have high linewidths compared to ``ordinary'' molecular cloud
structure.

For completeness, we verify that the differences are not due to
tracer: in the CMZ, cloud properties measured in dense gas tracers
\citep[N$_2$H$^+$, ][]{shetty12} fall in the same part of parameter
space as those measured by \citet{oka01} using \co{12}{1-0}.  In
30Dor-10, \co{12}{2-1} radii are 1.2$\pm$0.3 times the measured
\co{13}{2-1} radii, and linewidths are 1.05$\pm$0.15 times higher
-- neither shift affects the result shown here.  One might also
consider the thermal component of the line width -- parsec and smaller
clump linewidths are usually analyzed with the thermal width
removed, to test the hypothesis of hierarchical turbulent (nonthermal)
structure.  However, even at the elevated temperatures in 30Dor-10
(T$_{ex}\sim$40K, \S\ref{ltemass}), $\sigma_{v,th}$ is only 0.4$\;$\kms,
and a minor part of the linewidth.

Higher velocity dispersions than predicted by Larson's relation can
result if clouds are in virial equilibrium, but with external pressure
\citep[e.g.][]{field11}, in which case the virial equation is 
\[ {4\over 3}\pi r^3 P_{ext} = M\sigma_v^2 - {{GM^2}\over{5 r}}. \label{virial}\]
The pressure can be high inter-cloud thermal pressure, as invoked by
\citet{oka01} to explain the Milky Way central molecular zone.
However, the physical scales being measured in 30~Doradus are
significantly smaller.  A larger dispersion and a deviation of $<$10pc
clumps above the size-linewidth relation has been noted in many
Galactic clouds with different tracers
\citep[e.g.][]{heyer01,hennfalg}.  This deviation can be explained
either by clumps in quasi-equilibrium with interclump pressure,
i.e. the ``weight'' of the entire cloud acting on clumps within it
\citep{field11}, or equally well by clouds in chaotic gravitational
collapse, in which gravity drives turbulent motions and the
gravitational and kinetic energies remain balanced \citep{vs07}.  It
appears that the clumps in this extreme environment do not have
dramatically different properties from clumps of similar size in Milky
Way molecular clouds, or at least occupy the same part of parameter
space with similar dispersion.

\begin{figure}[h]
%\resizebox{!}{2.5in}{\includegraphics{mmol.mvir.eps}}
\resizebox{!}{3.5in}{\includegraphics{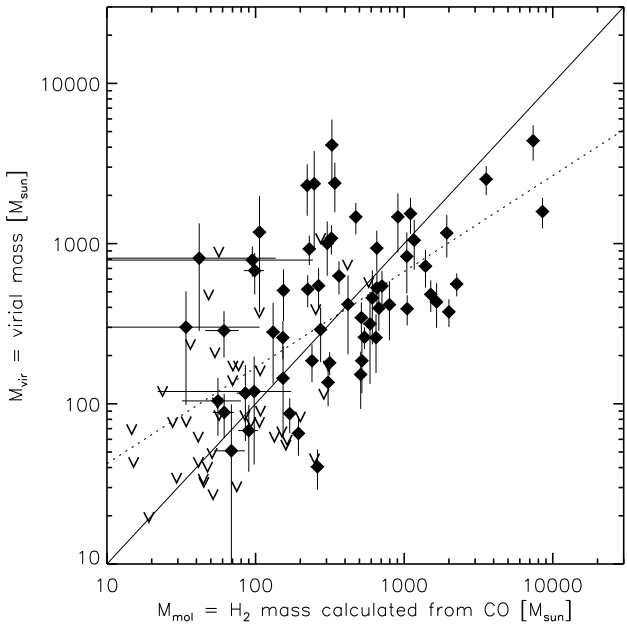}}
\caption{\label{mvirial} Molecular mass calculated from CO (M$_{mol}$) compared
  to virial mass.  The dotted line indicates a fit M$_{vir}$/M$_{mol}$
  =35$\;$M$_\odot\;$M$_{mol}^{-0.6}$.  Current observations are sensitive to
  M$\gtrsim$50$\;$M$_\odot$ depending on cloud size and excitation
  temperature.}
\end{figure}

The dynamic state of clumps and clouds can be further assessed by
directly comparing the gravitational and kinetic energies, or the
virial masses M$_{vir}$=5$\sigma_v^2r$/G and masses determined from CO
(M$_{mol}$); in 30Dor-10 these are in rough agreement
(Figure~\ref{mvirial}).
As discussed above, the dominant uncertainty in the molecular mass
calculated from CO is the $^{13}$CO abundance.  Calculating the
excitation temperature T$_{ex}$ from $^{12}$CO emission, assuming that
$^{12}$CO and $^{13}$CO have the same T$_{ex}$, and assuming that the
$^{13}$CO level population is described by a Boltzmann distribution at
T$_{ex}$ introduce another factor of $\sim$2 uncertainty
(\S\ref{ltemass}).
% SUB2
The \co{13}{2-1} sensitivity can be used to calculate the expected mass
sensitivity, although it is sensitive to the extitation temperature (equation
\ref{NCO}).  For a low \co{12}{2-1} brightness temperature of 5K, the
excitation temperature is 9.7K (eqn. \ref{Tex}). A clump with 5$\sigma$ peak
\co{13}{2-1} brightness of 0.75K, linewidth of 0.5$\;$km$\;$s$^{-1}$ (equal to
the velocity resolution), and a physical size equal to the minimum deconvolvable
size at that peak brightness \citep[see above and ][]{condon97}, has a mass of
20$\;$M$_\odot$.  A similarly bright clump the size of the beam has mass of
65$\;$M$_\odot$, consistent with the lowest points in subsequent figures.
%/SUB2
The dominant uncertainties in the virial mass are the assignment of a
``radius'' and use of the spherical virial theorem to a non-spherical
clump, and the assumption of virial equilibrium in the first place --
virial masses should be regarded as having at least a factor of 2
uncertainty.  Furthermore, in 30Dor-10, the selective dissociation of
CO relative to H$_2$ may make radii measured from CO emission, and
derived virial masses, {\em systematically} too small.

There is a trend for M$_{vir}$/M$_{mol}$ to decrease with increasing
M$_{mol}$.  The formal best-fit line is marked, M$_{vir}$/M$_{mol}$
=35$\;$M$_\odot\;$M$_{mol}^{-0.6}$.  \citet{bertoldi92} derive
M$_{vir}$/M$\propto$M$^{-2/3}$ for pressure-confined and magnetized
clumps.
The same decreasing trend M$_{vir}$/M$_{mol}\propto$M$_{mol}^\beta$ is
seen in dense clumps in many Galactic clouds, albeit often with a
shallower slope: Orion, $\beta$=-0.33 \citep{ikeda09}, Cygnus
\citep{dobashi96}, Cepheus and Casseopeia, $\beta$=-0.38
\citep{yonekura97}.  If the mass ratios in 30Dor-10 are due to
pressure confinement, the pressures required are high
P/k$\gtrsim$10$^6$.  The gravitational pressure from the weight of the
entire molecular cloud (inter-clump H$_2$ and H) can be estimated as
P$_G$/k
$\simeq$1.5$\;$cm$^{-3}\;$K$\;$(M$_{cloud}/M_\odot$)$^2\;$(r/pc)$^{-4}$
\citet{bertoldi92}. For 30Dor-10 P$_G$/k =
1.5(9$\times$10$^4$)$^2$(8)$^{-4}$ = 3$\times$10$^6$, with large
uncertainty due to the radius estimate, so the concept is not
unreasonable.  Analysis of the clump volume densities and inter-clump
gas physical conditions using {\it Herschel} spectroscopy of CII, OI,
OIII, and NII emission will help reveal the degree to which these
clumps are pressure-confined (Chevance et al. 2014, in prep).
% up to 1e7 or even 1e8

\begin{figure}[h]
%\resizebox{4in}{!}{\includegraphics{sizelinesigma.eps}}
\resizebox{4in}{!}{\includegraphics{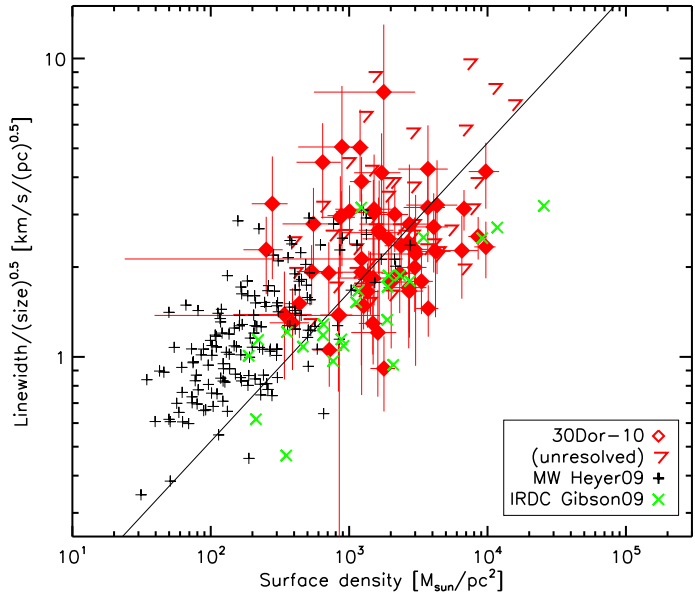}}
\caption{\label{sizelinesig} Clump properties in
  size-linewidth-surface density space.  Plus symbols are Milky Way
  clouds at $\sim$1-10pc scales analyzed with \co{13}{1-0} by Heyer et
  al. 2009.  Clumps in 30Dor-10 are marked with diamonds or sideways
  ``v'' symbols - the latter are clumps whose deconvolved radius is
  less than half of the synthesized beam, i.e. the radius used in the
  plot is an upper limit.  The line marks gravitational stability or
  virial equilibrium for spheres.  Clumps in 30Dor-10 extend to larger
  surface densities than the Milky Way clouds, as do dense clumps in
  Galactic IR-dark clouds (green crosses; data from Gibson et
  al. 2009)}
\end{figure}

Heyer et al. 2009 discuss the size-linewidth relation in detail and
find that the Milky Way relation is not merely a power-law but that
linewidth also depends on the cloud surface density.
Figure~\ref{sizelinesig} is a reproduction of their Figure~7, showing
clouds and clumps in size-linewidth-surface density parameter space.
The line marks the locus of gravitational stability or virial
equilibrium for spheres, namely $\sigma_v/\sqrt{r} = \sqrt{\pi G/5
  \Sigma}$.  Clouds following the Larson's scaling relations as well
as having a universal surface density would occupy a single point on
this plot.  Clumps in 30Dor-10 overlap in parameter space with Milky
Way clouds, but extend to higher surface densities for the same
linewidth-size parameter. One explanation for this population that
extends to higher surface density are that these clumps are not
supported by (thermal or nonthermal) motions against gravitational
collapse.  The most extreme (high surface density) clumps include
those half-dozen that have clearly associated protostars in near and
mid-infrared images. However, many high-surface density clumps also do
not obviously contain star formation in the infrared. It is possible
that these clumps are supported by magnetic fields, perhaps enhanced
by the material having been swept up by the massive stellar winds and
radiation.  Dense sub-parsec-sized clumps in infrared dark clouds in
the Milky Way display the same extension to larger surface densities,
as shown by the overplotted data on clumps in Galactic IR-dark clouds
\citep{gibson09}.

%=============================================================================
\subsection{Clump Masses:  Dust and CO Abundance and the X-factor}
\label{massmass}

\begin{figure}[h]
%\resizebox{!}{2.5in}{\includegraphics{dustmass.eps}}
\resizebox{!}{3in}{\includegraphics{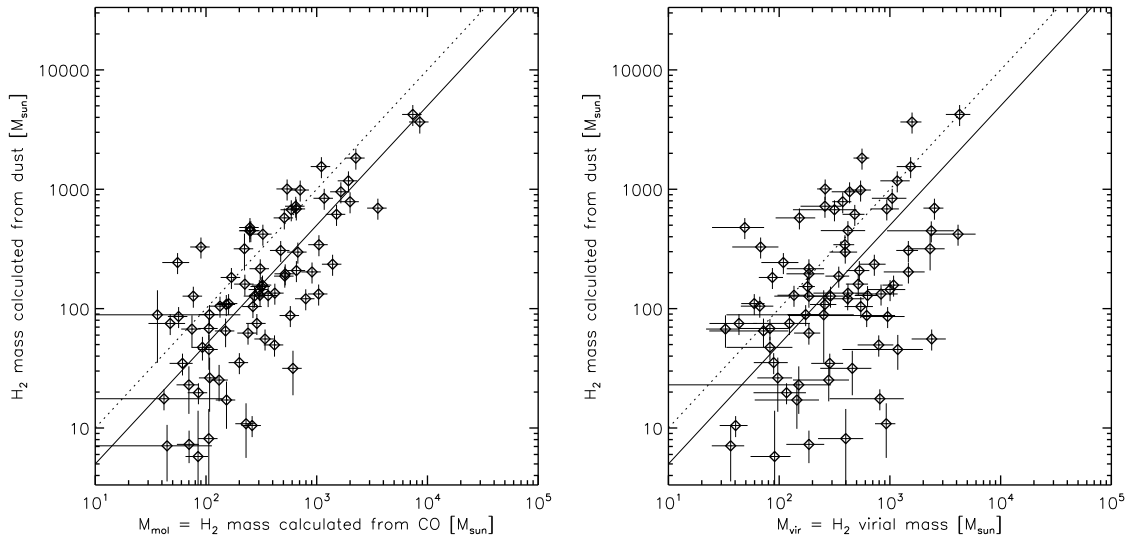}}
\caption{\label{dustmass} Comparison of clump molecular mass
  calculated from CO (M$_{mol}$, from dust continuum (M$_{dust}$), and
  using the virial theorem (M$_{vir}$).}
\end{figure}

Comparison of the masses of clumps calculated from dust (section
\ref{continuum}), using \co{12}{2-1} and \co{13}{2-1} (section
\ref{ltemass}), and from the virial theorem allows one to determine
the relative abundances of CO, dust, and H$_2$, and infer physical
conditions in the clumps. CO-derived and virial masses were compared
in the previous section; Figure~\ref{dustmass} compares these to the
dust-derived masses, showing that the dust masses are $\sim$2$\times$
lower than the virial or molecular masses.  This suggests that either
the gas-to-dust ratio in 30Dor-10 is twice the LMC average, or that
the dust temperatures are half the excitation temperatures measured
from $^{12}$CO.
Note that both the gas-to-dust ratio and the CO abundance
H$_2$/$^{13}$CO could be increased or decreased by the same factor
without changing the agreement between dust and CO-derived masses, but
a change by more than a factor $\sim$2 would then require explaining a
difference between dust-derived and virial mass. We do not find what
\citet{bot07} found in the SMC, virial masses exceeding dust-derived
masses and requiring additional cloud support.

The conversion ``X'' factor from CO emission to H$_2$ mass or column
density is used widely to measure molecular masses in galaxies at all
redshifts.  Conversion of optically thick emission (as most low-J
transitions of $^{12}$CO are) into mass requires that the molecular
clouds have relatively universal structure in space and velocity.
Even assuming that this is true, the conversion factor is expected to
depend on metallicity because H$_2$ is more effective at
self-shielding itself from photodissociation than CO.  At reduced
metallicity and consequently reduced dust abundance, ultraviolet
radiation penetrates more deeply into molecular clouds, and should
result in more of this selective photodissociation, and a larger ratio
of $H_2$ mass to CO emission.  The Milky Way X-factor has been
calculated with virial masses, dust emission, gamma-ray emission,
extinction, and optically thin isotopologue emission;
\citet{bolatto_araa} recommend
4.3$\pm$1.3$\;$M$_\odot\;$pc$^{-2}$(K$\;$km$\;$s$^{-1}$)$^{-1}$ or
2$\pm$0.6$\times$10$^{20}\;$cm$^{-2}\;$(K$\;$km$\;$s$^{-1}$)$^{-1}$
from the compilation.

Analysis of the $\sim$40pc resolution NANTEN \co{12}{1-0} survey of
the LMC suggested that the X-factor was higher than in the Milky Way:
19$\;$M$_\odot\;$pc$^{-2}\;$(K$\;$km$\;$s$^{-1}$)$^{-1}$
\citep{fukui99}, and this was corroborated by studies of selected
regions with SEST at $\sim$15pc resolution
\citep[28$\pm$4$\;$M$_\odot\;$pc$^{-2}\;$(K$\;$km$\;$s$^{-1}$)$^{-1}$][]{israel97}.
However, more recent analysis of the largely complete MAGMA survey
($\sim$15pc resolution) finds no strong evidence for a higher X-factor
in the LMC than in the Milky Way \citep{hughes10}.  It is
reasonable to expect the X-factor to be different in regions of
especially high radiation intensity, such as \dor.  \citet{pineda09}
investigated this with the same data as used by \citet{hughes10}, but
targeting specifically the $\sim$2kpc molecular cloud complex
extending south from \dor.  They found no variation of X-factor along
that molecular ridge.

\begin{figure}[h]
%\resizebox{!}{2.5in}{\includegraphics{xfactor.eps}}
\resizebox{!}{3.5in}{\includegraphics{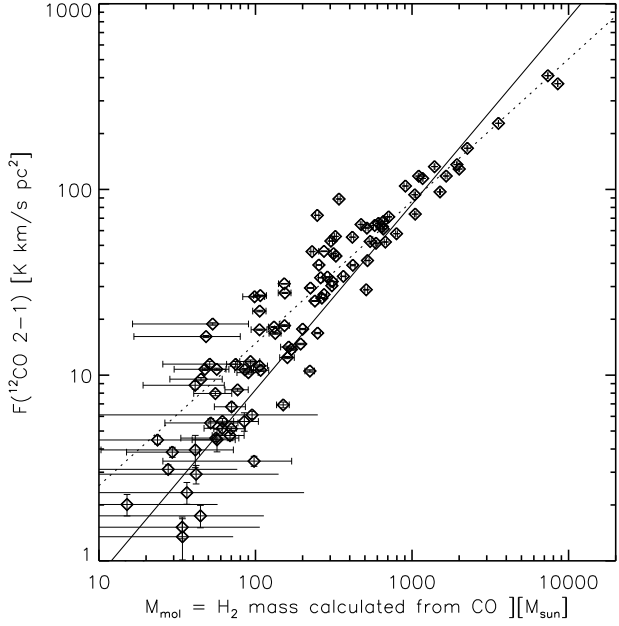}}
\caption{\label{xfactor} \co{12}{2-1} intensity integrated over each
  clump in space and velocity (F$_{12}$) as a function of mass
  calculated from CO (M$_{mol}$).  The solid line indicates the
  best-fit constant X-factor $\alpha$ = M$_{mol}$/F(\co{12}{2-1}) = 12$\pm$4$\;$M$_\odot\;$pc$^{-2}\;$(K$\;$km$\;$s$^{-1}$)$^{-1}$ and the dotted line a mass-dependent fit M$_{mol}$/F(\co{12}{2-1}) = 12(M/1300$\;$M$_\odot$)$^{0.23}$.
}
\end{figure}

We compare the calculated molecular mass (\S\ref{ltemass}) to the
\co{12}{2-1} intensity and calculate M$_{mol}$/F(\co{12}{2-1}) =
12$\pm$4$\;$M$_\odot\;$pc$^{-2}\;$(K$\;$km$\;$s$^{-1}$)$^{-1}$
(Fig.~\ref{xfactor}).  Applying the same ratio
F(\co{12}{1-0})/F(\co{12}{2-1})=0.7 as \citet{sandstrom} yields an
X-factor of
8.4$\pm$3$\;$M$_\odot\;$pc$^{-2}\;$(\co{12}{1-0}$\;$K$\;$km$\;$s$^{-1}$)$^{-1}$
{\em for clumps} in 30Dor-10.  This is remarkably almost exactly the
Milky Way value divided by the LMC metallicity.  There is a possible
trend of lower X-factor at higher mass, namely
M$_{mol}$/F(\co{12}{2-1}) = 12(M/1300$\;$M$_\odot$)$^{0.23}$.  This
could be explained by the more massive clumps having a higher beam
filling fraction.  Alternately, as discussed in \S~\ref{ltemass}, the
excitation temperature derived from $^{12}$CO may be higher than the
actual $^{13}$CO excitation temperature in bright clumps, and our
calculated mass too high for those clumps.

Figure~\ref{xfactor} only considers CO emission associated with dense
clumps, but we can also perform the calculation for the entire map:
the total calculated H$_2$ mass {\em traced by CO} in 30Dor-10 is
8.9$\pm$0.3$\times$10$^4\;$M$_\odot$, and the total \co{12}{2-1} flux
9400$\;$K$\,$km$\,$s$^{-1}\;$pc$^2$, which translates into an overall
M$_{mol}$/F(\co{12}{2-1})
9.5$\;$M$_\odot\;$pc$^{-2}\;$(K$\;$km$\;$s$^{-1}$)$^{-1}$ --
consistent with the factor derived from clumps and with the fact that
most \co{12}{2-1} emission is associated with clumps.

This method of calculating a mass using an optically thin isotopologue
only measures H$_2$ associated with CO, and not ``CO-dark'' H$_2$.
The X-factor including any H$_2$ that may exist between and not
associated with clumps is expected to be higher, but we cannot measure
that directly.  One could in principle calculate the total gas mass
from the total dust emission, and derive a total X-factor from that,
but unfortunately we do not have the single-dish data to correct for
large-scale dust emission resolved out by the interferometer.  The
most direct comparison to our data would be to an X-factor derived for
resolved molecular clouds using the optically thin isotopologue
method; indeed \citet{heyer01} perform this calculation for the Milky
Way (albeit with poorer physical resolution than our data in
30Dor-10), and derive approximately a factor of two lower X-factor;
\citet{duval} perform a similar calculation with different cloud
definitions and find a value consistent with
4.3$\pm$1.3$\;$M$_\odot\;$pc$^{-2}$(K$\;$km$\;$s$^{-1}$)$^{-1}$.  We
conclude that the X-factor in dense clumps in 30Dor-10 is consistent
with merely being scaled by metallicity, may be up to a factor of two
higher depending on the Milky Way comparison value chosen, and that a
measurement of any inter-clump ``CO-dark'' H$_2$ is likely to yield a
significantly higher value.

%=============================================================================
\subsection{Clump Mass Distribution}

\begin{figure}
%\resizebox{2.5in}{!}{\includegraphics{mhist.eps}}
\resizebox{3in}{!}{\includegraphics{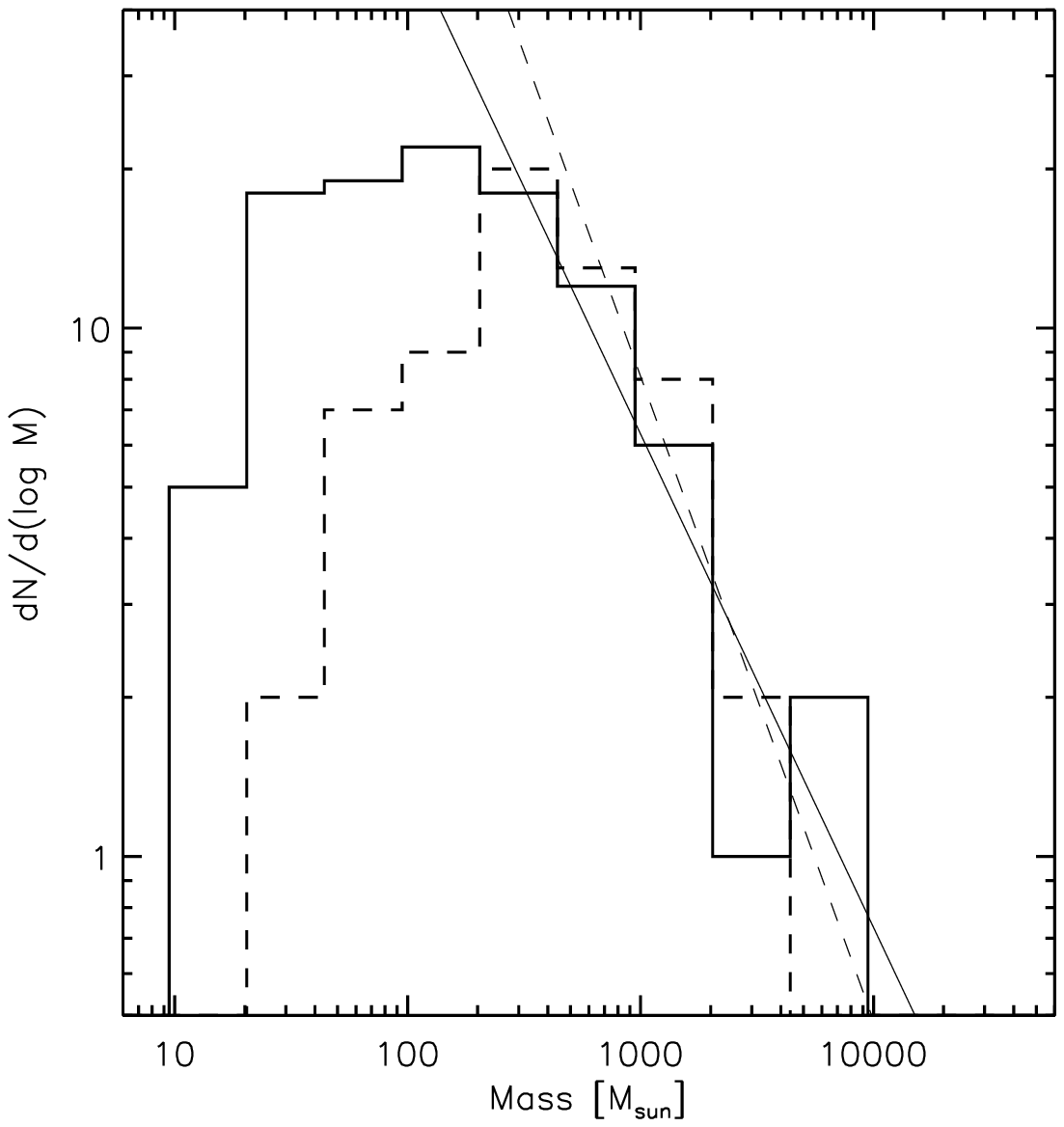}}
\caption{\label{mhist} Clump mass histogram in 30Dor-10. The solid line shows molecular masses calculated from $^{12}$CO and \co{13}{2-1}, the dashed line shows virial masses. The best-fit slopes dN/d(log M) $\propto$ M$^\alpha$ for M$>$500M$_\odot$ are $\alpha$=-0.9$\pm$0.2 and -1.2$\pm$0.2 for molecular and virial masses, respectively.}
\end{figure}

We conclude our discussion of clump masses with the clump mass
distribution (Fig.~\ref{mhist}).  The clump decomposition tends to
select regions similar to the beamsize, and such mass histograms are
fraught with caveats \citep{peril}.  Nevertheless, the clumps in 30Dor-10
exhibit similar behaviour to those in Galactic clumps on similar scales
\citep[e.g. Orion;][]{ikeda09}.  The best-fit slopes dN/d(log M)
$\propto$ M$^\alpha$ for M$>$500M$_\odot$ are $\alpha$=-0.9$\pm$0.2
and 1.2$\pm$0.2 for molecular and virial masses, respectively.
Galactic clump distributions have slopes $\simeq$-1 (-1.1 in Orion).
As noted above, M$_{virial}$/M$_{mol}$ decreases with increasing mass,
causing the slope of the virial mass distribution to be steeper.
% SUB2
Although higher resolution observations will break up the clumps observed here, the slope of the mass distribution is not dramatically sensitive to that, as shown by varying-resolution simulations in \citet{reid10}.
% /SUB2

%=============================================================================
\subsection{Trends with Position (or Lack Thereof)}

\begin{figure}
%\centerline{\resizebox{3in}{!}{\includegraphics{trend_x.eps}}
%\resizebox{3in}{!}{\includegraphics{trend_m.eps}}}
%\centerline{\resizebox{3in}{!}{\includegraphics{trend_r.eps}}
%\resizebox{3in}{!}{\includegraphics{trend_sep.eps}}}
\centerline{\resizebox{3in}{!}{\includegraphics{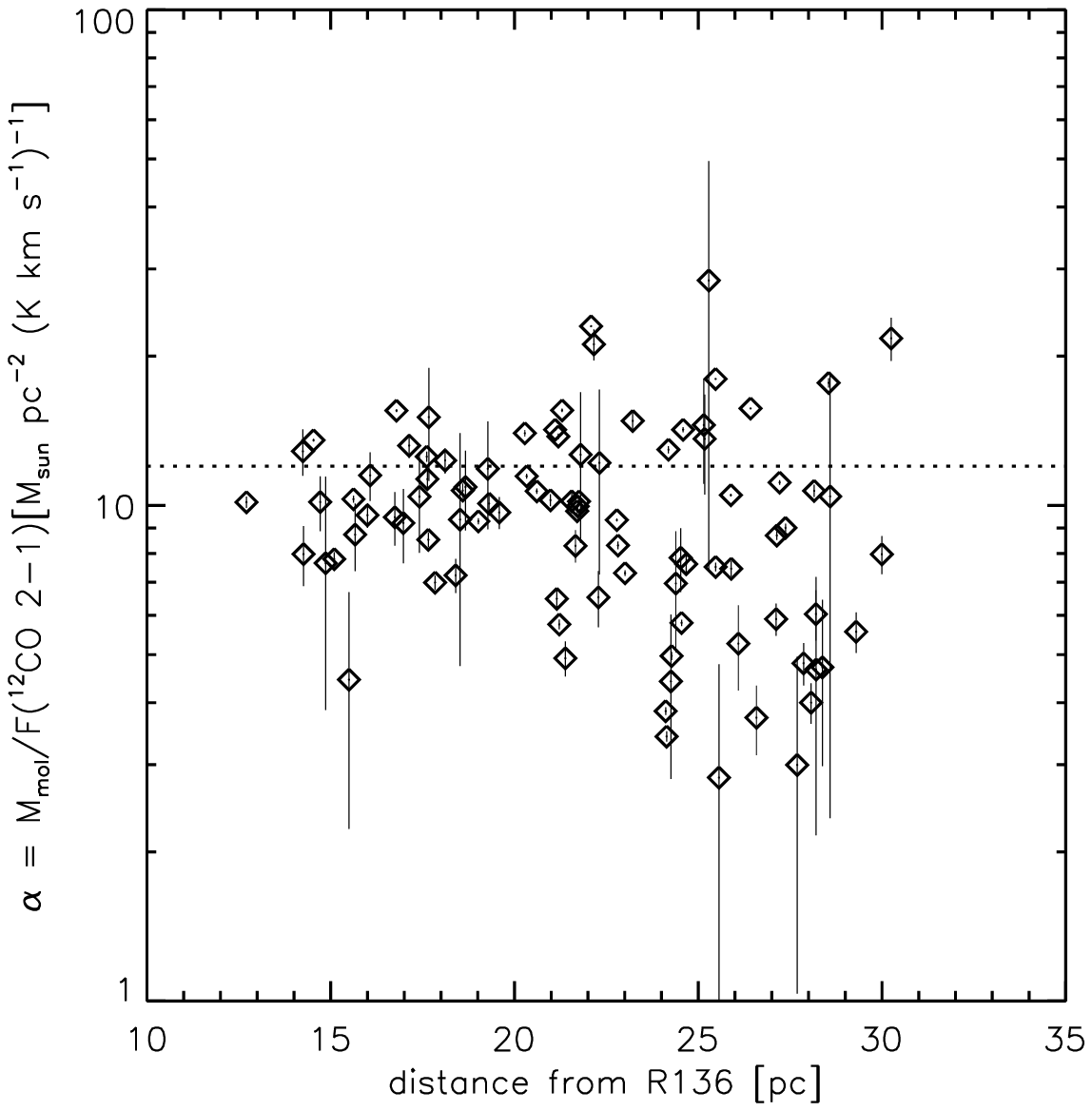}}
\resizebox{3in}{!}{\includegraphics{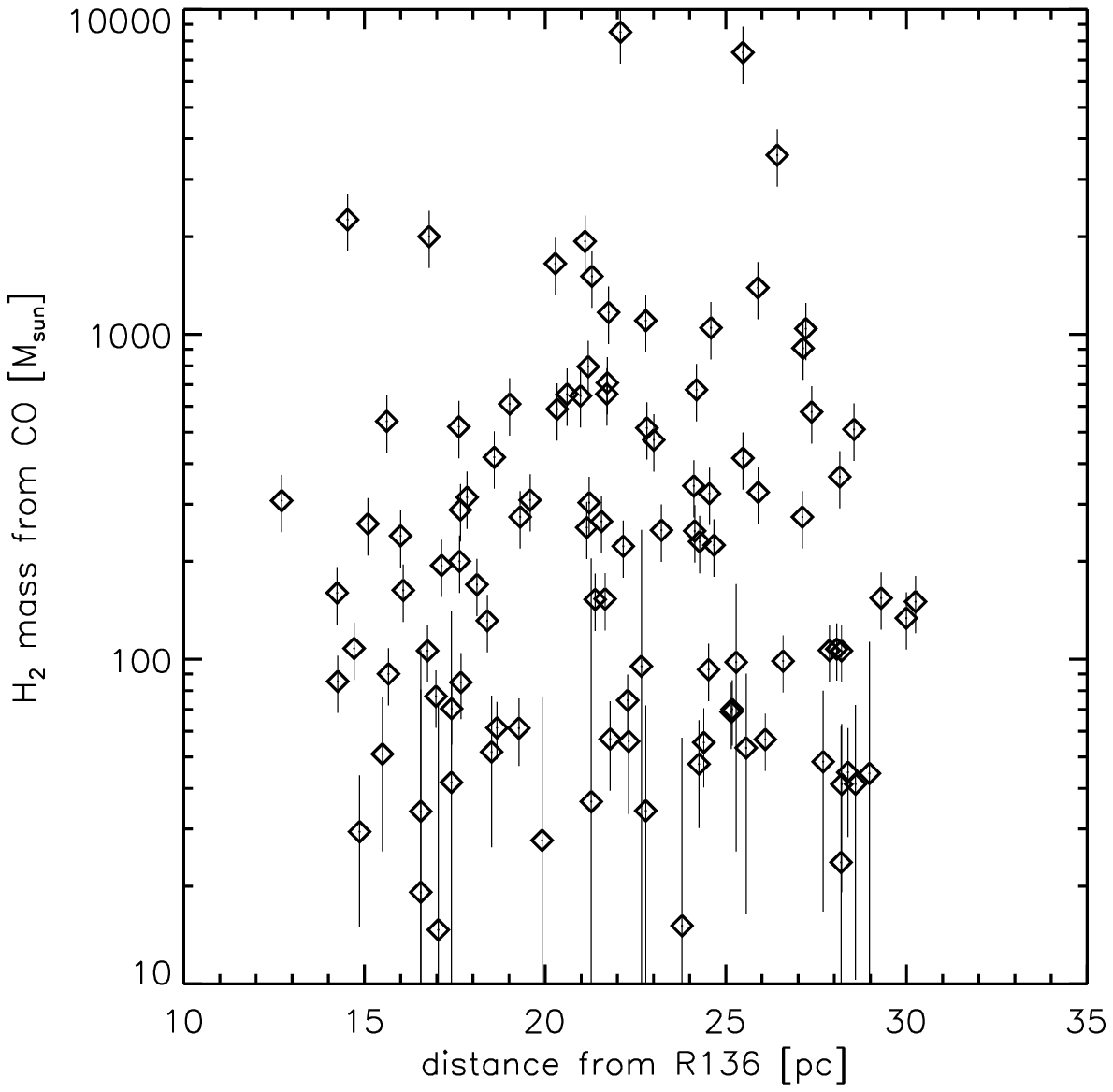}}}
\centerline{\resizebox{3in}{!}{\includegraphics{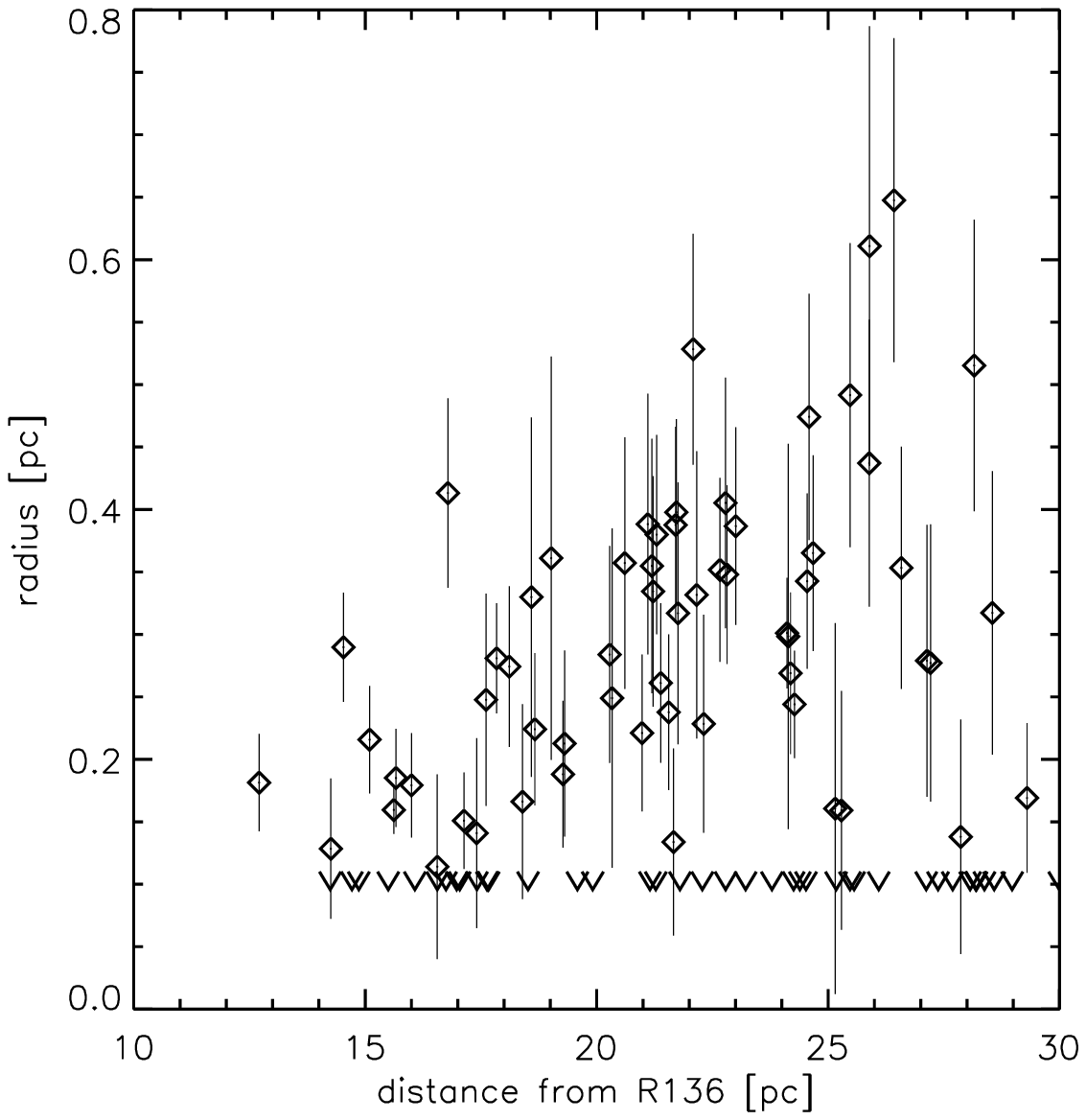}}
\resizebox{3in}{!}{\includegraphics{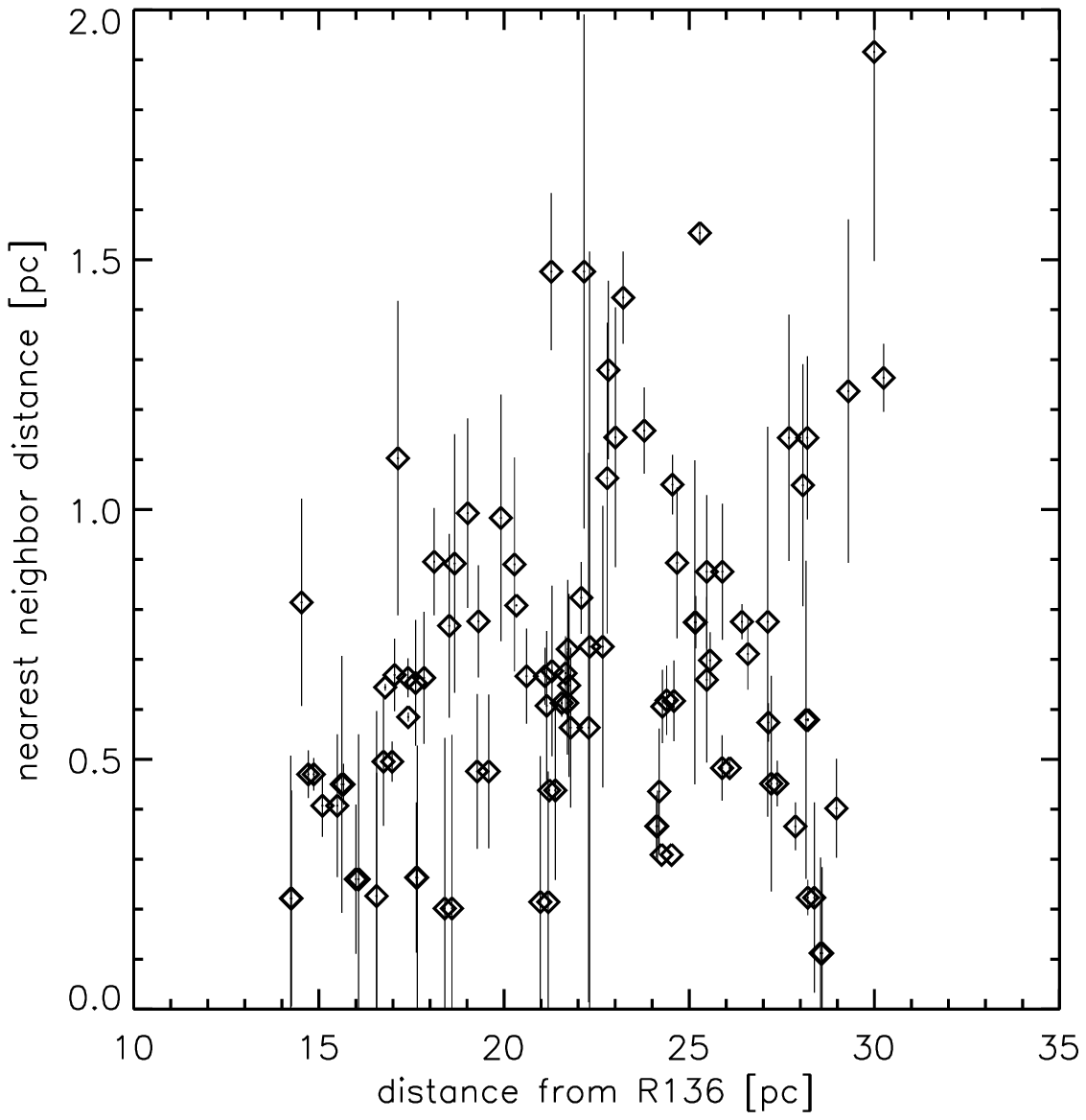}}}
\caption{\label{trends} Trends in clump properties as a function of
  distance from R136.  The properties of dense clumps do not vary
  strongly as a function of depth into the molecular cloud.  Two weak
  trends are that the average size of resolved clumps increases, and
  the average separation of clumps increases, i.e. the volume fraction
  of the cloud contained in clumps decreases. }
\end{figure}

If the intense radiation field from R136 is affecting the molecular
cloud, one might expect to see trends of clump properties with
increasing distance from R136, going deeper into the cloud (clump
projected distances from R136 are $\sim$13--30pc.)
Figure~\ref{trends} shows that this is not obviously the case, except
for a tendency for the clumps to be more widely separated deeper in
the cloud -- i.e. the volume fraction of the cloud that is in dense
clumps is somewhat less deeper into the cloud, perhaps due to a
different spatial spectrum of turbulence or less fragmentation.  The
uncertainty in separation plotted here is the difference between the
nearest neighbor distance, and 1/$\sqrt{2}$ times the distance to the
second nearest neighbor.  There is also a weak trend for the size of
the resolved clumps to increase with distance away from R136, although
there is no lack of unresolved clumps further back in the cloud as
well.

The lack of strong trends in clump properties is consistent with the
overall conclusion that dense clumps in this molecular cloud are not
significantly different from dense clumps in other environments, but
that the larger-scale cloud physics may be affected.  In this case,
the cloud may be compressed overall closer to the PDR front, bringing
the dense clumps closer together, but once dense clumps form, their
properties do not apparently vary strongly with depth into the cloud.
If this result holds generally, it would support a ``normal'' stellar
initial mass function in super-star clusters: In 30~Doradus the
super-star cluster is already formed, and current star formation will
not amount to another super-star cluster, but we do have evidence that
lower-mass stars and small clusters form ``normally'' in dense clumps
even in the vicinity of many massive stars.  It is not well-understood
whether low-mass stars form before or after massive ones in SSC
formation, but our data supports the robustness of the process.

It is also important to remember the 3-dimensional nature of the
cloud. The wispy blue NIR emission (Fig.~\ref{nir_ysos}) coincident
with some of the CO features suggests that radiation from R136 may
reach structures that are in projection towards the top of our mapped
area, but in fact are in front or behind the bulk of molecular
material.  If there are trends of clump properties with distance into
the cloud, the 3D geometry likely disguises them.

%==============================================
\section{Summary}
\label{conclusion}

The giant molecular cloud 30Dor-10, north of R136 in 30~Doradus, was
observed with ALMA during Early Science Cycle~0 and with APEX.  This cloud
is strongly irradiated and has lower than solar metal abundances, so
is an ideal laboratory in which to test whether either of metallicity or incident radiation field can affect the physical conditions of molecular gas.
The primary resulting data are image cubes of \co{12}{2-1},
\co{13}{2-1}, C$^{18}$O$\;$2-1, and an image of the 1.3$\;$mm continuum.
Analysis of these data leads to the conclusions about clumpy
molecular cloud structure listed below.  

\begin{enumerate}
\item CO emission in 30~Doradus is highly structured -- the molecular
  cloud is composed of dense clumps and filaments, very similar to
  Galactic molecular clouds.  The relative contribution of lower
  column density inter-clump regions to the total $^{12}$CO flux is
  however smaller than in Galactic clouds like W3. This suggests that there
  is significant photodissociation of CO by radiation penetrating
  between the dense clumps.  For the first time in an extreme
  environment like 30~Doradus, we observe directly the decrease in
  $^{12}$CO emission relative to total gas mass that manifests
  as an increased X-factor in more distant unresolved molecular
  clouds.
% SUB2
\item We decompose these data into discrete structures or clumps, to
  analyze trends in physical properties.  This procedure is sensitive
  to only a modest range of physical scales, and higher resolution
  observations will reveal substructure, but it is instructive to
  compare to Milky Way clouds decomposed at similar physical
  resolution.
% /SUB2
  Dense clumps measured for the first time on sub-parsec scales
  outside of our Galaxy have properties not unlike those in Galactic
  giant molecular clouds.  The clumps have somewhat larger velocity
  dispersions than predicted by the size-linewidth relation for Milky
  Way clouds.  One interpretation is that the clumps in 30Dor-10 are
  pressure-confined; the ratio of virial mass (M$_{vir}$) to mass
  calculated from CO emission (M$_{mol}$) decreases with increasing
  M$_{mol}$ as predicted for pressure-confined clumps.  However, such
  a decrease is seen in many Galactic clouds on small scales, and the
  interpretation is not unique.  Selective photodissiciation of CO
  relative to H$_2$ may make the measured sizes of clumps measured
  from CO emission, and calculated virial masses, too small.
\item In size-linewidth-surface density space, clumps in 30Dor-10
  overlap with the locus populated by $>$10pc Galactic clouds but extend to
  higher surface densities at the same size and linewidth.  This trend
  agrees with sub-parsec clumps in Galactic infrared-dark clouds
  (IRDCs).
\item Masses calculated from dust M$_{dust}$ are approximately half of
  those calculated from CO.  This could be explained by a gas-to-dust
  ratio half of the LMC average, or could be a result of assuming that
  the dust temperature equals the CO excitation temperature calculated
  from the \co{12}{2-1} brightness temperature.  \co{12}{2-1} is quite
  optically thick, thus its excitation temperature in the emitting clump
  surface may be warmer than the temperature of most of the dust;
  assuming too high of a dust temperature would result in low dust
  masses.  However, the dust masses are also a factor of two lower
  than virial masses, which suggests that the reduced dust-to-gas
  ratio may be the more likely explanation.  One can increase or
  decrease both the assumed gas-to-dust and H$_2$/$^{13}$CO ratios and
  not change the ratio of calculated dust mass to mass calculated from
  CO, but changing either of those abundances dramatically would
  result in inconsistency with the virial masses.  We suggest that the
  the CO abundance {\em in dense clumps} in 30~Doradus is similar to
  the outer Milky Way H$_2$/$^{13}$CO=5$\times$10$^6$, and the
  gas-to-dust ratio is $\sim$600.
  Average abundances for the entire molecular cloud including
  inter-clump regions, may be different. 
\item Use of optically thick $^{12}$CO emission to trace mass {\em
    within molecular clouds} is not generally advised, but we
  calculate that for dense clumps $\alpha$ = M$_{mol}$/F(\co{12}{2-1})
  = 12$\pm$4$\;$M$_\odot\;$pc$^{-2}\;$(K$\;$km$\;$s$^{-1}$)$^{-1}$.
  The data suggest a decrease in $\alpha$ in brighter, more massive
  clumps.  The measured X-factor is 2 times the average value in the
  Milky Way and other solar-metallicity galaxies, again merely the
  scaling expected for the LMC metallicity.  Since most of the
  $^{12}$CO emission is associated with dense clumps, The ratio of
  total cloud mass to $^{12}$CO emission may be higher when
  inter-clump H$_2$ is included.
\item The high end of the dense clump mass function has a power-law
  slope of -1.0$\pm$0.2, similar to sub-parsec sized clumps in
  Galactic massive star formation regions.
\item Clump physical properties are not a strong function of distance
  from the dominant source of radiation, the R136 cluster core.  In
  particular, we measure no trend in the X-factor or clump mass with
  distance from R136.  There is a weak trend for the clumps to be
  larger and further from each other as one travels back into the
  molecular cloud away from R136.  In other words the fraction of the
  molecular cloud filled with dense clumps is larger closer to the PDR
  edge where radiative feedback is strongest.
\end{enumerate}

ALMA is beginning to dissect star formation regions and star-forming
molecular gas on previously inaccessible size scales.  For the first
time we can study molecular cloud physics resolving structures smaller
than 1pc outside the Milky Way.  Studying physics like this in a wide
range of environments promises to revolutionize our understanding of
star-forming molecular clouds.

\acknowledgements{
We thank the APEX staff for their support during the observations.  We
thank the ALMA staff whose hard work made these observations possible
-- obtaining science-grade data from an array still being commissioned
was a remarkable accomplishment.
The National Radio Astronomy Observatory is a facility of the National Science
 Foundation operated under cooperative agreement by Associated Universities, Inc.
This paper makes use of the following ALMA data: ADS/JAO.ALMA\#2011.0.00273.S. ALMA is a partnership of ESO (representing its member states), NSF (USA) and NINS (Japan), together with NRC (Canada) and NSC and ASIAA (Taiwan), in cooperation with the Republic of Chile. The Joint ALMA Observatory is operated by ESO, AUI/NRAO and NAOJ.
This research made use of Montage, funded by the National Aeronautics
and Space Administration's Earth Science Technology Office,
Computation Technologies Project, under Cooperative Agreement Number
NCC5-626 between NASA and the California Institute of
Technology. Montage is maintained by the NASA/IPAC Infrared Science
Archive.
M. Meixner was supported by NASA NAG5-12595.
T.V. was supported by a National Science Foundation astrochemistry grant to the University of Virginia.
}

\clearpage
%========================================================

\end{document}